\documentclass[10pt,conference]{IEEEtran}
\usepackage{amsmath,amssymb,amsfonts}
\usepackage{algorithmic}
\usepackage{graphicx}
\usepackage{textcomp}
\usepackage{xcolor}
\usepackage[hyphens]{url}
\usepackage{multirow}
\usepackage{booktabs}
\usepackage{pifont}
\usepackage{array}
\usepackage{cite}
\usepackage{tikz}

\usepackage{fancyhdr}

\usepackage{hyperref}

\usepackage{xcolor}
\usepackage{colortbl}
\usepackage{xspace}
\usepackage{soul}

\newcommand{\sect}[1]{Section~{\ref{#1}}}

\newcommand{\eqn}[1]{Equation~{\ref{#1}}}
\newcommand{\eqns}[1]{Equations~{\ref{#1}}}
\newcommand{\fig}[1]{Figure~{\ref{#1}}}

\newcommand{\tab}[1]{Table~{\ref{#1}}}

\def\naive{na\"{\i}ve\xspace}

\makeatletter
\DeclareRobustCommand\onedot{\futurelet\@let@token\@onedot}
\def\@onedot{\ifx\@let@token.\else.\null\fi\xspace}

\def\eg{\emph{e.g}\onedot} 
\def\ie{\emph{i.e}\onedot}

\makeatother

\definecolor{mydarkblue}{rgb}{0,0.08,1}
\definecolor{mydarkgreen}{rgb}{0.02,0.6,0.02}
\definecolor{mydarkred}{rgb}{0.8,0.02,0.02}
\definecolor{mydarkorange}{rgb}{0.40,0.2,0.02}
\definecolor{mypurple}{rgb}{111,0,255}
\definecolor{myred}{rgb}{1.0,0.0,0.0}
\definecolor{mygold}{rgb}{0.75,0.6,0.12}
\definecolor{myblue}{rgb}{0,0.2,0.8}
\definecolor{mydarkgray}{rgb}{0.66,0.66,0.66}

\def\adg{ADG\xspace}
\def\dag{DAG\xspace}

\def\matrixIDmath{\mathbf{M}_{I\rightarrow D}}
\def\matrixTImath{\mathbf{M}_{T \rightarrow I}}
\def\matrixSImath{\mathbf{M}_{S \rightarrow I}}

\def\matrixID{$\mathbf{M}_{I\rightarrow D}$\xspace}
\def\matrixTI{$\mathbf{M}_{T \rightarrow I}$\xspace}
\def\matrixSI{$\mathbf{M}_{S \rightarrow I}$\xspace}

\newcommand{\cmark}{\ding{51}}%
\newcommand{\xmark}{\ding{55}}%

\def\frontend{front end\xspace}

\newcommand{\redcircle}[1]{
    \tikz[baseline=(char.base)]{
        \node[shape=circle,fill=red,draw=red,text=white,inner sep=0.5pt, font=\small] (char) {#1};
    }
}

\newcommand{\blackcircle}[1]{
    \tikz[baseline=(char.base)]{
        \node[shape=circle,fill=black,draw=black,text=white,inner sep=0.5pt, font=\small] (char) {#1};
    }
}

\newcommand{\hpcayear}{2025}

\newcommand{\hpcasubmissionnumber}{1067}
\title{LEGO: Spatial Accelerator Generation and Optimization for Tensor Applications}

\def\hpcacameraready{} 

\newcommand\hpcaauthors{Yujun Lin$^{*}$, Zhekai Zhang$^{*}$, Song Han $\thanks{Thanks}$}
\newcommand\hpcaaffiliation{Massachusetts Institute of Technology}
\newcommand\hpcaemail{\{yujunlin,zhangzk,songhan\}@mit.edu}

\author{
  \ifdefined\hpcacameraready
    \IEEEauthorblockN{\hpcaauthors{}}
      \IEEEauthorblockA{
        \hpcaaffiliation{} \\
        \hpcaemail{} \\
        \urlstyle{tt}
        {\footnotesize \textcolor{blue}{\url{https://hanlab.mit.edu/projects/lego}}}
      }
  \else
    \IEEEauthorblockN{\normalsize{HPCA \hpcayear{} Submission
      \textbf{\#\hpcasubmissionnumber{}}} \\
      \IEEEauthorblockA{
        Confidential Draft \\
        Do NOT Distribute!!
      }
    }
  \fi 
}

\fancypagestyle{camerareadyfirstpage}{%
  \fancyhead{}
  
  \fancyhead[C]{
    \ifdefined\aeopen
    \parbox[][12mm][t]{13.5cm}{\hpcayear{} IEEE International Symposium on High-Performance Computer Architecture (HPCA)}    
    \else
      \ifdefined\aereviewed
      \parbox[][12mm][t]{13.5cm}{\hpcayear{} IEEE International Symposium on High-Performance Computer Architecture (HPCA)}
      \else
      \ifdefined\aereproduced
      \parbox[][12mm][t]{13.5cm}{\hpcayear{} IEEE International Symposium on High-Performance Computer Architecture (HPCA)}
      \else
      \parbox[][0mm][t]{13.5cm}{\hpcayear{} IEEE International Symposium on High-Performance Computer Architecture (HPCA)}
    \fi 
    \fi 
    \fi 
    \ifdefined\aeopen 
      \includegraphics[width=12mm,height=12mm]{ae-badges/open-research-objects.pdf}
    \fi 
    \ifdefined\aereviewed
      \includegraphics[width=12mm,height=12mm]{ae-badges/research-objects-reviewed.pdf}
    \fi 
    \ifdefined\aereproduced
      \includegraphics[width=12mm,height=12mm]{ae-badges/results-reproduced.pdf}
    \fi
  }
  \fancyfoot[C]{}
}
\begin{document}
\maketitle

\ifdefined\hpcacameraready 
  \thispagestyle{camerareadyfirstpage}
  \pagestyle{empty}
\else
  \thispagestyle{plain}
  \pagestyle{plain}
\fi

\newcommand{\hpcaheight}{0mm}
\ifdefined\eaopen
\renewcommand{\hpcaheight}{12mm}
\fi

\def\thefootnote{*}\footnotetext{Equal Contributions.}\def\thefootnote{\arabic{footnote}}

\begin{abstract}
Modern tensor applications, especially foundation models and generative AI applications require multiple input modalities (both vision and language), which increases the demand for flexible accelerator architecture. 
Existing frameworks suffer from the trade-off between design flexibility and productivity of RTL generation: either limited to very few hand-written templates or cannot automatically generate the RTL.

To address this challenge, we propose the LEGO framework, which targets tensor applications and automatically generates spatial architecture design and outputs synthesizable RTL code without handwritten RTL design templates.
Leveraging the affine-transformation-based architecture representation,
LEGO front end finds interconnections between function units, synthesizes the memory system, and fuses different spatial dataflow designs based on data reuse analysis.
LEGO back end then translates the hardware in a primitive-level graph to perform lower-level optimizations, and applies a set of linear-programming algorithms to optimally insert pipeline registers and reduce the overhead of unused logic when switching spatial dataflows.

Our evaluation demonstrates that LEGO can achieve 3.2 speedup and 2.4× energy efficiency compared to previous work Gemmini, and can generate one architecture for diverse modern foundation models in generative AI applications.

\ifdefined\arxiv
Project page: \url{https://hanlab.mit.edu/projects/lego}
\else
\fi
\end{abstract}

\section{Introduction}
\label{sect:introduction}
The proliferation of tensor applications, particularly deep neural networks, has led to an unprecedented demand for efficient and high-performing solutions. 
Especially in the era of multi-modal large foundation models, a single accelerator needs to process/generate both language and vision inputs/outputs~\cite{chatgpt,touvron2023llama,copilot}, such as GPT-4~\cite{gpt4} and stable diffusion~\cite{ho2020denoising}.
Tensor operations play a vital role in these applications, and thus many specialized spatial hardware architectures have been proposed, such as TPU~\cite{jouppi2017datacenter} for general matrix multiplication (GEMM), Eyeriss~\cite{chen2016eyeriss} for convolutional neural networks, and A3~\cite{ham2020a3}, SpAtten~\cite{wang2021spatten}, and Sanger~\cite{lu2021sanger} for attention operation in transformers~\cite{vaswani2017attention}, and PointAcc~\cite{lin2021pointacc} for sparse convolution in point cloud deep learning~\cite{choy20194d}. 
Although these architectures have exhibited remarkable performance for the target operations, their development cycle is frequently impeded by extensive human effort. This drawback puts them at a disadvantage when handling emerging algorithms that require diverse new operations.
Moreover, these human-designed accelerators cannot cover the entire design space of architectures, making them sub-optimal for certain situations. Therefore, an automatic hardware design methodology is urgently needed to reduce human effort and shorten the development cycle for highly flexible architectures.

\begin{figure}[t]
    \centering
    \includegraphics[width=\linewidth]{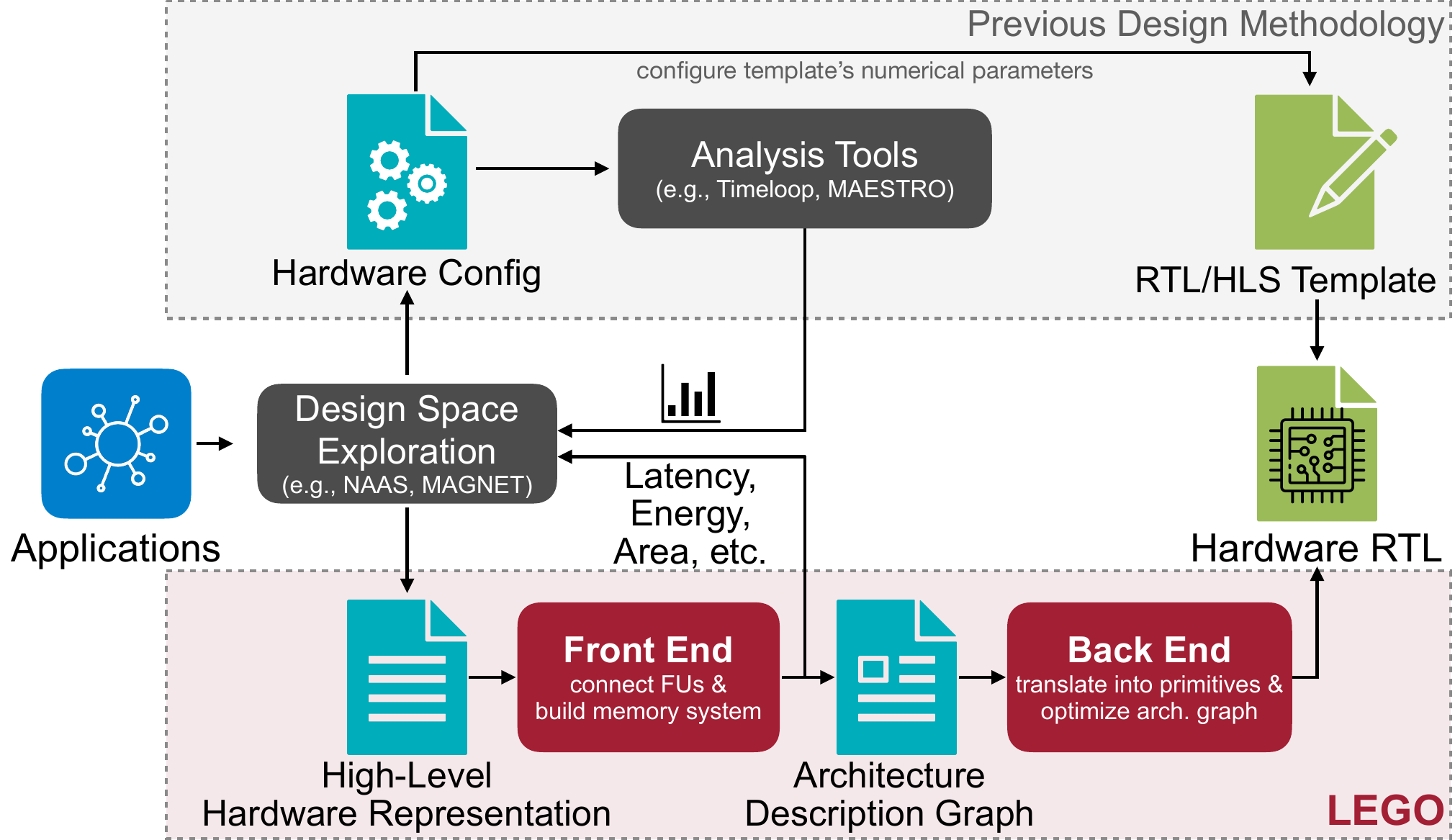}
    \caption{Instead of configuring the sizing parameters in the hardware template, LEGO directly generates spatial architecture design and outputs RTL code from high-level hardware description.}
    \label{fig:methodology}
    \vspace{-12pt}
\end{figure}

In support of spatial architecture development for tensor operations in the applications, dataflow analysis tools like Timeloop~\cite{parashar2019timeloop}, MAESTRO~\cite{kwon2019understanding}, and TENET~\cite{lu2021tenet} have been proposed to model the hardware behavior and predict their latency and energy. Recently, design space exploration tools such as NAAS~\cite{lin2021naas} and MAGNET~\cite{venkatesan2019magnet} have leveraged these tools to search for optimal architectures in the design space.

Aside from dataflow analysis frameworks that are not capable of generating the actual hardware, there are also recent works that focus on RTL generation. \cite{genc2021gemmini,venkatesan2019magnet,sharma2016dnnweaver,zhang2020dna} rely on handwritten RTL/HLS code of accelerator designs with numerical sizing parameters, which is rigid and limits the hardware design space, ultimately impeding their practical usefulness. 
AutoSA~\cite{wang2021autosa} and Tensorlib~\cite{jia2021tensorlib} further get rid of the handwritten architecture template and extend the design space to any 2D array.
However, these frameworks only support hardware generation for a single tensor kernel, making it difficult to accommodate the increasing diversity of tensor kernels used in a single application. Different tensor kernels may benefit from distinct dataflows to optimize performance, necessitating the fusion of multiple dataflows in a single hardware implementation.
Another challenge in the prior frameworks is the intertwining of control flow (e.g., wave-like propagation in a systolic array) with the dataflow representation. This entanglement complicates the analysis of data reuse, introducing division and modulo operations in the representation and prompting previous work to only capture a single direction of reuse for simplification. It also inhibits the reuse analysis of the control signals (\eg, memory address), resulting in the control logic (\eg, address generator) with 2.0{×} area overhead and 2.6{×} power overhead.

\begin{table}[t]
\centering
\caption{Comparison between Different Spatial Accelerator Analysis and Generation Frameworks}
\renewcommand{\arraystretch}{1.2}
\scalebox{0.8}{
\footnotesize
\begin{tabular}{
m{70pt}|
>{\centering\arraybackslash}m{70pt}
>{\centering\arraybackslash}m{70pt}}
\toprule
{\centering \textbf{Work}}
 & {\textbf{RTL Generation}} 
 & {\textbf{Design Space}}
 \\\midrule
Timeloop~\cite{parashar2019timeloop}, MAESTRO~\cite{kwon2019understanding},
TENET~\cite{lu2021tenet}
   & \xmark
   & any dataflows\\\hline
Gemmini~\cite{genc2021gemmini}, DNNWeaver~\cite{sharma2016dnnweaver}
   & \cmark{ }{ }{ }{ }(Template-based)
   & limited to { }{ }{ }{ }{ }{ }{ }{ }{ }{ }{ }{ }{ }{ }{ } 2 templates\\ \hline
MAGNET~\cite{venkatesan2019magnet}
   & \cmark{ }{ }{ }{ }(Template-based)
   & limited to { }{ }{ }{ }{ }{ }{ }{ }{ }{ }{ }{ }{ }{ } 1 template\\ \hline
DNA~\cite{zhang2020dna}
   & \cmark{ }{ }{ }{ }{ }{ }{ }{ }{ }{ }{ }{ }{ }{ }{ }{ }{ }(Template-based)
   & limited to { }{ }{ }{ }{ }{ }{ }{ }{ }{ }{ }{ }{ }{ } 3 dataflows\\ \hline
AutoSA~\cite{wang2021autosa}
   & \cmark{ }{ }{ }{ }{ }{ }{ }{ }{ }{ }{ }{ }{ }{ }{ }{ }{ }{ }{ }{ }{ }(HLS for FPGA)
   & limited to { }{ }{ }{ }{ }{ }{ }{ }{ }{ } 2D systolic array\\ \hline
Tensorlib~\cite{jia2021tensorlib}
   & \cmark
   & limited to { }{ }{ }{ }{ }{ }{ }{ }{ }{ } 2D array\\ \hline
SODA+MLIR+Bambu\cite{agostini2022soda,lattner2021mlir,ferrandi2021bambu}
   & \cmark{ }{ }{ }{ }{ }{ }{ }{ }{ }{ }{ }{ }{ }{ }{ }{ }{ }{ }{ }{ }{ }(HLS-based)
   & any dataflows\\ \hline
LEGO (ours)
   & \cmark
   & any dataflows and their combinations\\ \bottomrule
\end{tabular}
}
\vspace{-9pt}
\label{tab:comparison}
\end{table}

To address these challenges, we propose a novel framework, LEGO, as shown in \fig{fig:methodology}, targeting the tensor operations in the applications such as matrix multiplication, tensor factorization, and decomposition. LEGO directly generates spatial architecture design and outputs synthesizable RTL code from a comprehensive high-level hardware abstraction. 

\begin{figure*}[t]
    \centering
    \includegraphics[width=\linewidth]{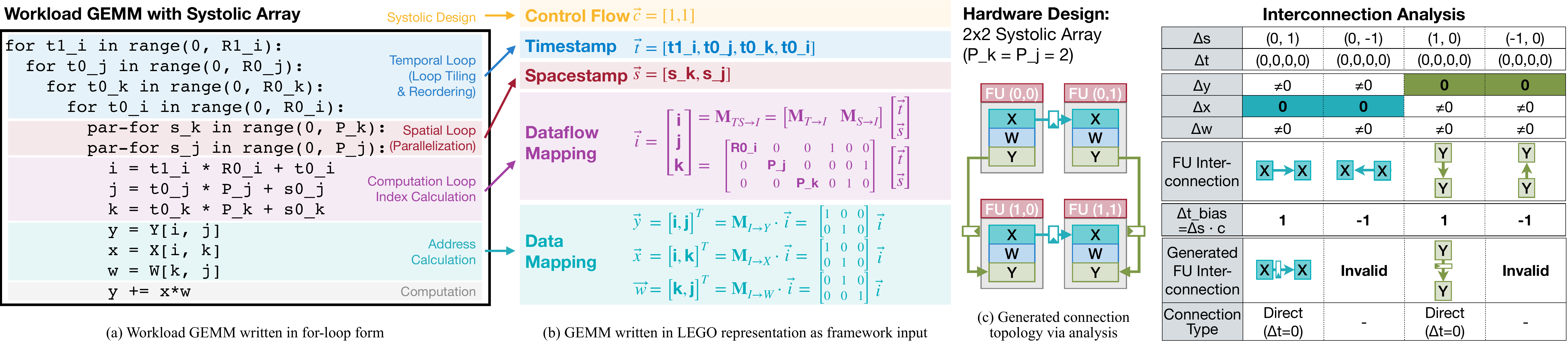}
    \vspace{-18pt}
    \caption{Example of GEMM parallelizing \texttt{j} and \texttt{k} dimensions (TPU~\cite{jouppi2017datacenter} arch): (a) GEMM in the conventional for-loop form; (b)
    LEGO uses affine transformations to represent GEMM; (c) LEGO generates direct FU interconnections ($\Delta \vec{s}$) for every possible data reuse ($\Delta \vec{x} = 0$, $\Delta \vec{y} = 0$).}
    \label{fig:gemm-notation-example}
\end{figure*}

Specifically, the input of LEGO system is the tensor workload that can be written in the form of loop nests as shown in Figure~{\ref{fig:gemm-notation-example}}. LEGO first decouples the generation of computation logic, data paths, and memory system, enabling full design freedom in interconnection topology. The essence of connection generation is to take full advantage of data reuse, and thus LEGO employs a relation-centric representation entirely based on linear affine transformations, thereby simplifying the data reuse analysis. LEGO also enables the control signal reuse by separating the control flow from the dataflow representation, reducing the control logic overhead.

LEGO front end captures the data behavior in the target tensor operation, formulates the connectivity decision into a graph optimization problem, and determines the functional units (FUs) connectivity and memory allocation. It also enables to fuse connections for various dataflows in a single design, which allows the execution of end-to-end tensor applications while minimizing the data path overhead and improving the energy efficiency by up to 20\% compared to {\naive} design fusion with multiplexers.

The introduction of higher freedom in connection topology and dataflow fusion leads to much more complexity in the structure of the design. Unlike prior work that uses manually designed FU array templates to construct the hardware, which can be difficult to cover all possible situations, and can be suboptimal due to the lack of global information, we propose LEGO back end as a systematic way to optimize the FU array as a whole with multiple transformation passes on a lower level representation, resulting in 35\% area savings and 28\% area savings over {\naive} RTL generation without optimization.

LEGO emerges as a versatile tool for the design of specialized tensor cores within general-purpose platforms. Its adaptability and design efficiency make it an ideal candidate for enhancing the computational power and energy efficiency of heterogeneous systems.
\begin{figure*}[t]
    \centering
    \includegraphics[width=\linewidth]{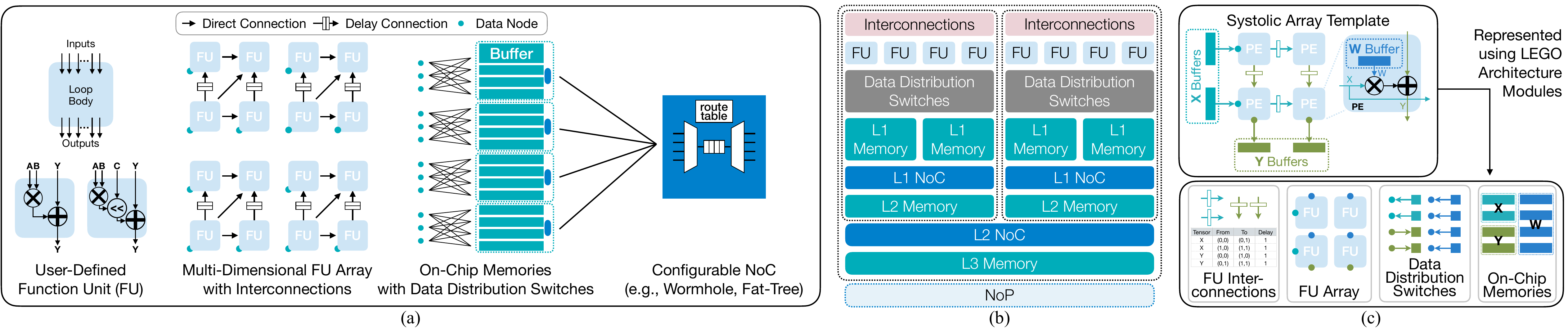}
    \vspace{-20pt}
    \caption{(a) LEGO Spatial Architecture decouples computation logic, data path topology, and memory system. (b) LEGO modules are assembled hierarchically. (c) An example of representing a conventional systolic array template using LEGO modules.  }
    \label{fig:architecture}
    \vspace{-9pt}
\end{figure*}
In summary, the contributions of this work are as follows:
\begin{itemize}
    \item Provides an automatic solution, LEGO framework, that streamlines the hardware development process from tensor kernels in the applications to RTL code based on hierarchical LEGO spatial architecture (\sect{sect:overview}).
    \item Introduces a high-level hardware representation based on the mappings between the iteration domain and (par)for-loop instances (\sect{sect:representation}).
    \item Builds an FU-level spatial architecture graph by solving integer linear equations and optimizes the graph via an MST-based algorithm and a BFS-based heuristic algorithm to minimize the data path overhead (\sect{sect:frontend}).
    \item Optimizes the primitive-level architecture graph via a set of linear-programming-based algorithms to reduce register cost and unused connection overhead (\sect{sect:backend}).
    \item LEGO outperforms Gemmini~\cite{genc2021gemmini} with 3.2× speedup and 2.4× energy efficiency.
\end{itemize}

\section{LEGO Spatial Architecture Design Space}
\label{sect:overview}
Different dataflows have different memory access patterns, leading to different connection topologies in the spatial architecture.
However, as illustrated by the systolic array template in Figure~{\ref{fig:architecture}}(c), the handwritten RTL templates often integrate the compute logic and memories as a single entity, and fix the connections between processing elements (PEs) and buffers~{{\cite{wang2021autosa, jia2021tensorlib, jia2022ems, genc2021gemmini, venkatesan2019magnet, zhang2020dna,sharma2016dnnweaver}}}. Thereby, the template-based generation frameworks are either numerical sizing or limited to a few predefined topology choices.

Noting that the essence of generating a spatial accelerator for any given dataflow is the creation of the corresponding connection topology, LEGO decouples the design of the computational units, data paths, and memory system, as depicted in Figure~{\ref{fig:architecture}}(a). Subsequently, these modules are reassembled in a hierarchical manner to formulate an integrated accelerator design, as illustrated in Figure~{\ref{fig:architecture}}(b). Data paths between functional units (FUs) are managed by the module ``FU Interconnections'', and paths between FUs and on-chip memories are organized into a separate module ``Data Distribution Switches''. By generating these two modules, LEGO is able to handle the RTL generation for any given dataflow and dataflow combinations. Specifically, the basic modules in LEGO include:

\textbf{Functional Units (FUs)} execute the computation defined in the loop body. For example, the FU of GEMM and CONV is multiplication-add (MAC): $Y += A \cdot B$; the FU of the mixed-precision GEMM using BitFusion~\cite{sharma2018bit} is a 2-bit mult-shift-add unit: $Y += (A \cdot B) << C$. LEGO can employ a user-defined FU design to offer workload customization.

\textbf{FU Interconnections} specify the data communication between functional units, \eg, 4 store-and-forward paths in Figure~{\ref{fig:architecture}}(c). 
These connections are automatically inferred and optimized in LEGO during hardware generation time.
There are two types of FU interconnections: direct and delay interconnection. Delay interconnections are essentially FIFOs
whose depths are programmable during runtime, offering a configurable delay when propagating the data. 
If two FUs are connected with multiple data paths, based on the dataflow configuration, only one path is activated at every cycle during runtime, forming an acyclic forest to enforce the direction of data transfer.
This gets rid of the complex scheduling and prevents any potential deadlock.

\textbf{Data Distribution Switches} route the data between on-chip memories and functional units, which are generated by LEGO during hardware generation time. It can be direct connections in Figure~{\ref{fig:architecture}}(c), a chubby tree in MAERI~\cite{kwon2018maeri} and multi-cast in Eyeriss~\cite{chen2016eyeriss}. As the data reuse between FUs is handled by FU interconnections, data distribution switches resolve the memory access conflicts due to the tensor data layout in on-chip memories. This allows a simple memory system design without complicated analysis in previous work EMS{~\cite{jia2022ems}}.

\textbf{Memories} are parameterized by their capacity and width. The shape of the memory array is determined by LEGO. 
L1 memories are assigned to different tensors. Each L1 memory space has only one address generator and controller.
L2 and higher-level memories are shared by all tensors and use a Buffet-based interface to buffer the transmitted data.

\textbf{Network-on-Chip (NoC)} connects different memories at the same level and interacts with those at the higher level. Since data layout in L1 memories is determined by the data access pattern in the FU array while data layout in higher-level memories is in line with the layout in off-chip memories, L1 NoC supports strided memory access and tensor transpose in addition to communication. LEGO currently supports multiple predefined NoC structures, including the multi-stage butterfly network and wormhole NoC. The deadlock is prevented using a classical X-Y routing method for wormhole NoC.

\textbf{Post-processing units (PPUs)} process the computations that cannot trivially be mapped to the FU array. For example, activation functions such as softmax and normalization functions in neural networks.
Each post-processing unit consists of a lookup table to calculate the activation function and a reduction unit to summarize the mean/variance during the normalization or sum of exponents during softmax. 
The post-processing units share the output buffers with the FU array to support in-place calculations and minimize hardware overhead.

\textbf{Scalability.} LEGO architecture embodies a hierarchical structure that offers flexibility for scaling up the design. At the lowest level, the number of FUs can be increased to directly boost computational capacity and throughput. Alternatively, at the higher level, the design accommodates the expansion of Processing Elements (PEs) interconnected via the L2 NoC without changing PE microarchitecture. 

\begin{figure*}[t]
    \centering
    \includegraphics[width=\linewidth]{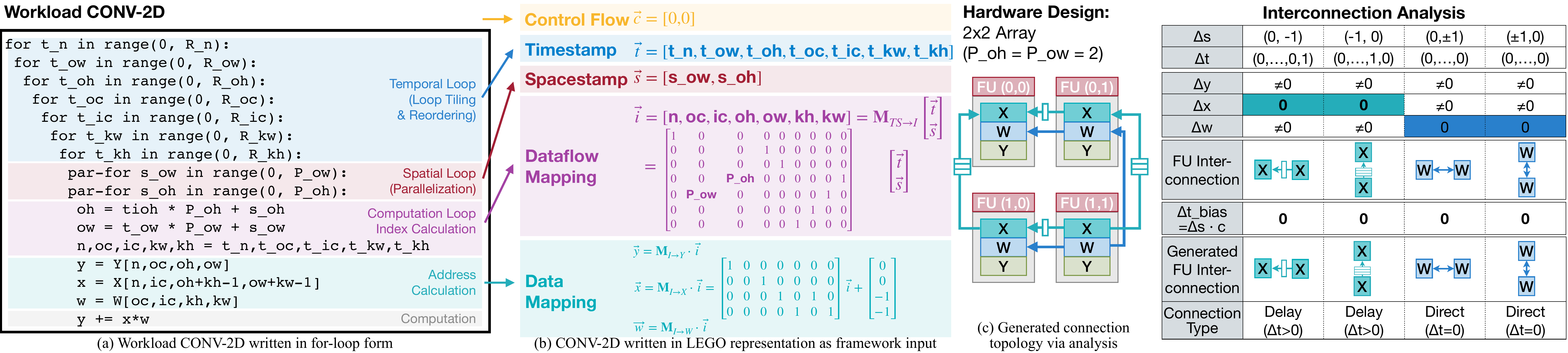}
    \vspace{-18pt}
    \caption{Example of 2D Convolution parallelizing \texttt{oh} and \texttt{ow} dimensions (ShiDianNao~\cite{du2015shidiannao} architecture): (a) Conv2D in the conventional for-loop form; (b) LEGO uses affine transformations to represent Conv2D; (c) LEGO generates FU interconnections ($\Delta \vec{s}$) for every possible data reuse ($\Delta \vec{x} = 0$, $\Delta \vec{w} = 0$).}
    \label{fig:conv-notation-example}
    \vspace{-9pt}
\end{figure*}

\section{LEGO Inputs: Relation-Centric Representation}
\label{sect:representation}

The key mechanism of data path generation in LEGO lies in maximizing the reuse: find the patterns of when a tensor element is accessed in the FU and how it is reused across FUs, and finally, determine the interconnections between FUs and the allocation of data on memories. As shown in the left of Figure~{\ref{fig:gemm-notation-example}} and~{\ref{fig:conv-notation-example}}, a tensor operation is essentially the computation in the form of loop nests, which are composed of the following parts:
\begin{itemize}
    \item the temporal iteration index $\vec{t}$, which represents the for-loop instance (\ie, temporal dataflow). As in the blue box, $\vec{t} = [{t_1}_i, {t_0}_j, {t_0}_k, {t_0}_i]$ for GEMM and $\vec{t}=[t_{n}, t_{ow}, t_{oh}, t_{oc}, t_{ic}, t_{kw}, t_{kh}]$ for Conv2D. Dimensions of time domain $T$ may not be independent due to loop tiling.
    \item the spatial iteration index $\vec{s}$, which represents the parfor-loop instance (\ie, spatial dataflow). As in the red box, $\vec{s} = [s_k, s_j]$ for GEMM and $\vec{s} = [s_{ow}, s_{oh}]$ for Conv2D.
    \item the computation iteration index $\vec{i}$, which represents the naive loop index before any tiling, that is, independent dimensions of computation iteration domain $I$.  As in the purple box, $\vec{i}=[i, j, k]$ for GEMM and $\vec{i}=[n, oc, ic, oh, ow, kh, kw]$ for Conv2D.
    \item the tensor data indexes $\vec{d}$. As in the green box, the index of output tensor Y is $\vec{y}=[i,j]$, and that of input tensor X is $\vec{x}=[i,k]$ for GEMM.
    \item the computation definition. As in the green box, the computation of both GEMM and Conv2D is $Y += X \cdot W$.
    \item the direction and delay of the control signals (\eg, cycle number, memory addresses). As in the yellow box, the control signals are propagated along $s_j$ and $s_k$ dimensions with 1 cycle delay (\ie, in the systolic manner), and thus control signal flow $\vec{c}=[1, 1]$.
\end{itemize}

The hardware is clocked by the temporal index ($\vec{t}$), FU is identified by the spatial index ($\vec{s}$), and the data accessed (\ie, memory address) are determined by the data indexes (\eg, $\vec{y}$). Hence, the relation between these indexes is crucial for analyzing the data reuse across FUs. LEGO exploit the computation iteration index $\vec{i}$ as as a key intermediary:
\begin{itemize}
    \item the relation $f_{I\rightarrow D}$ between the computation iteration index $\vec{i}$ and tensor data indexes (\eg, $\vec{y}$, $\vec{x}$) is \textit{hardware-agnostic} and defines the fundamental data access characteristics of the workload ({\ref{sect:workload-representation}}).
    \item the relation $f_{TS\rightarrow I}$ between temporal/spatial iteration index ($[\vec{t}\; \vec{s}]$) and computation iteration index $\vec{i}$ is \textit{workload-agnostic} and defines how the loop nests are tiled, reordered and parallelized on the spatial accelerator ({\ref{sect:dataflow-representation}}).
\end{itemize}

These two relations ($f_{I\rightarrow D}$ and $f_{TS\rightarrow I}$) along with the computation definition and the control signal flow ($\vec{c}$) are the input of LEGO framework. 

\subsection{Workload Representation}
\label{sect:workload-representation}
In LEGO, the characteristics of a tensor workload, such as GEMM or Conv2D, are represented by (a) computation in the loop body, which is defined using the primitive discussed later in LEGO back end, and (b) data mapping:

\noindent\textit{Definition 1:} \textbf{Data Mapping.} Given a $n_I$-dimensional computation iteration domain $I$ and a $n_D$-dimensional tensor domain $D$, a data mapping relation is defined as,
\begin{equation}
    \vec{d} = f_{I\rightarrow D}(\vec{i}) = \mathbf{M}_{I\rightarrow D}\vec{i} + \mathbf{b}_{I\rightarrow D},
    \label{eqn:data-mapping}
\end{equation}
where $\vec{d}$ is a $n_D$-d vector indicating the index of an element in tensor $D$, $\vec{i}$ is a $n_I$-d vector indicating a computation iteration index in $I$, and \matrixID is a affine transformation matrix with a size of $n_D \times n_I$, $\mathbf{b}_{I\rightarrow D}$ is a $n_D$-d bias.

\fig{fig:gemm-notation-example} shows an example of the GEMM workload $\mathbf{Y}_{i,j}=\mathbf{X}_{i,k}\mathbf{W}_{k,j}$ ($\vec{y}=[i,j]$, $\vec{x}=[i,k]$, $\vec{w}=[k,j]$ ) where
\begin{equation*}\tiny
\vec{y}=\begin{bmatrix}
1 & 0 & 0\\ 
0 & 1 & 0 \\
0 & 0 & 0
\end{bmatrix}\begin{bmatrix}
i\\ 
j\\ 
k
\end{bmatrix}, \vec{x}=\begin{bmatrix}
1 & 0 & 0\\ 
0 & 0 & 0 \\
0 & 0 & 1
\end{bmatrix}\begin{bmatrix}
i\\ 
j\\ 
k
\end{bmatrix}, \vec{w}=\begin{bmatrix}
0 & 0 & 0\\ 
0 & 0 & 1 \\
0 & 1 & 0
\end{bmatrix}\begin{bmatrix}
i\\ 
j\\ 
k
\end{bmatrix}
\end{equation*}
These three 3-by-2 matrices compose the relation $f_{I\rightarrow D}$, serving as one of the inputs to the LEGO framework.

\fig{fig:conv-notation-example} shows another example of 2D-Conv workload $\mathbf{Y}_{n, oc, oh, ow}=Conv\left(\mathbf{X}_{n, ic, ih, iw}, \mathbf{W}_{oc, ic, kh, kw}\right)$. The three 4-by-7 matrices ($\mathbf{M}_{I\rightarrow Y}$, $\mathbf{M}_{I\rightarrow X}$, $\mathbf{M}_{I\rightarrow W}$) in the green box make up the the relation $f_{I\rightarrow D}$.

\subsection{Dataflow Representation}
\label{sect:dataflow-representation}

In LEGO, loop ordering, tiling, and parallelization are represented by the dataflow mapping:

\noindent\textit{Definition 2:} \textbf{Dataflow Mapping}. Given a $n_I$-dimensional computation iteration domain $I$, $n_T$-level for-loop nests $T$, $n_S$-level parfor-loop nests $S$, the dataflow mapping is,
\begin{equation}
    \vec{i}  = f_{TS\rightarrow I}(\vec{t}, \vec{s}) = \begin{bmatrix}\mathbf{M}_{T \rightarrow I}~~\mathbf{M}_{S \rightarrow I} \end{bmatrix}\begin{bmatrix}\vec{t}\\ \vec{s}\end{bmatrix}.
    \label{eqn:dataflow}
\end{equation}
$\vec{s}$ is a $n_S$-d vector indicating the FU coordinates in a $n_S$-d FU array. $\vec{t}$ is a $n_T$-d vector indicating the for-loop state index. The lexicographical order of $\vec{t}$ indicates the for-loop execution order. That is, the first dimension of $\vec{t}$ is the outermost for-loop and the last dimension of $\vec{t}$ is the innermost for-loop. Therefore, given the for-loop sizes $\vec{R_T}$, we can easily convert $\vec{t}$ into a scalar integer timestamp $t$ as follows,
\begin{equation} \label{eqn:time-scalar}
    t= \begin{bmatrix}0&\vec{t}\;^T\end{bmatrix}\begin{bmatrix}\vec{R_T}\\1\end{bmatrix} = \left(\left(t_0 \cdot R_{T_1} + t_1\right) \cdot R_{T_2} + t_2\right)\cdots
\end{equation}
where $\vec{t}=[t_0, t_1, \cdots]^T$ and $\vec{R_T} = [R_{T_0}, R_{T_1}, R_{T_2}, \cdots]^T$.

One can infer the \matrixTI and \matrixSI from the for-loop sizes $\vec{R_T}$ and parfor-loop sizes $\vec{R_S}$ (\ie, the FU array sizes), since they only contain values from the loop sizes.
In the GEMM example of \fig{fig:gemm-notation-example}(purple region), we have for-loop sizes $\vec{R_T}=[R_{1_i}, R_{0_j},R_{0_k},R_{0_i}]$ and parfor-loop sizes $\vec{R_S}=[P_k, P_j]$. Given the for-loop iteration index $\vec{t}=[t_{1_i}, t_{0_j},t_{0_k},t_{0_i}]$ and the parfor-loop iteration index $\vec{s}=[s_k, s_j]$, the dataflow mapping can be inferred from the purple area:
\begin{equation*}\tiny
    \vec{i}=\begin{bmatrix}i\\j\\k\end{bmatrix} = \begin{bmatrix}
R_{0_i} & 0   & 0   & 1 & 0 & 0 \\
0       & P_j & 0   & 0 & 0 & 1 \\
0       & 0   & P_k & 0 & 1 & 0 \\
\end{bmatrix}\begin{bmatrix}t_{1_i}\\t_{0_j}\\t_{0_k}\\t_{0_i}\\s_k\\s_j\end{bmatrix}
\end{equation*}
This 3-by-6 matrix is relation $f_{TS\rightarrow I}$, serving as another input to the LEGO framework.
\fig{fig:conv-notation-example}(purple region) shows another example of expressing 2D convolution dataflow in ShiDianNao~\cite{du2015shidiannao} architecture.

\subsection{Control Flow Representation.}

In addition to tensor data, different FUs may share the same control signals, including data valid/invalid signal and memory buffer addresses. 
Therefore, LEGO introduces control flow representation, denoted by a $n_S$-d vector $\vec{c}$. 
Sharing the control signals can be implemented as a store-and-forward logic: a positive (or negative) value in $\vec{c}$ indicates the signals are propagated forward (or backward) along the corresponding spatial dimension with a delay. Value zero means signals are directly broadcast to the FUs along the dimension.

In the example of \fig{fig:gemm-notation-example}, since control signals are forwarded along both spatial dimensions with one cycle delay, the control flow is $\vec{c}=[1,1]$. While in \fig{fig:conv-notation-example}, all FUs share the control signals at the same time, and thus $\vec{c}=[0,0]$. 

Since the delay of control signals results in the delay of timestamp $t$, a timestamp bias $t_{bias}$ can be assigned to each FU to represent such delay. $t_{bias}$ can be calculated as,
\begin{equation}
    t_{bias, \vec{s}} = \vec{s}\;^T \cdot\vec{c}
\end{equation}
and the difference in timestamp $\vec{t}$ of any two FUs ($\vec{s}_1$ and $\vec{s}_2$) can be calculated as,
\begin{equation}
    {\Delta t}_{bias, \vec{s}_1 \rightarrow \vec{s}_2} = (\vec{s}_2 - \vec{s}_1)^T\cdot\vec{c} = {\Delta \vec{s}}^T\cdot\vec{c}
\end{equation}

\subsection{Differences to Previous Representations}
\label{sect:representation-differences}

Previous work, including polyhedral model{~\cite{wang2021autosa}} and STT{~\cite{baltus1993efficient,jia2021tensorlib,jia2022ems}}, tend to represent the dataflow as a mapping from the computation iteration domain $I$ to the temporal domain $T$ and spatial domains $S$. For example, GEMM in Figure~{\ref{fig:gemm-notation-example}} is represented as,

\vspace{-5pt}
{\footnotesize
\begin{align*}
    I[i,j,k] \rightarrow \{ & S[j\;\%\;P_j, k\;\%\;P_k]~| \\
    & T[j\;//\;P_j, k\;//\;P_k, j\;\%\;P_j + k\;\%\;P_k + i] \}
\end{align*}
}
In contrast, LEGO dataflow representation maps from temporal and spatial indexes to the iteration domain to eliminate division and modulo operations.

Furthermore, previous representations did not consider the data flow of control signals and took the timestamp $\vec{t}$ as a global absolute time for all FUs $\vec{s}$ in the data mapping representation. That is $t=5$ means cycle 5, and $j\;\%\;P_j + k\;\%\;P_k$ has to be added to the temporal indexes to represent a systolic array design. Therefore, previous work will generate an individual control logic (such as counters and address generators) for each FU and memory buffer, leading to a huge consumption of area and power. 
Instead, LEGO decouples the control signals from the dataflow representation and views timestamp $\vec{t}$ as a local execution time for each FU $\vec{s}$. LEGO treats the systolic manner of sharing tensor data as a natural result of the systolic manner of sharing control signals. 
As in the example of Figure~{\ref{fig:gemm-notation-example}}, since control flow is $\vec{c}=[1,1]$, multicast/reduction-tree is converted to systolic forward/reduction by adding a delay of $\Delta t_{bias}=1$ on corresponding FU interconnections. Therefore, LEGO only generates one control unit and propagates the control signals among FUs according to the control flow vector $\vec{c}$, leading to 6.5{×} flip-flop savings and 5.0{×} LUT savings compared to polyhedral-based AutoSA{~\cite{wang2021autosa}}, and 2.0{×} area savings and 2.6{×} power savings compared to STT-based TensorLib{~\cite{jia2021tensorlib}}.
\section{LEGO Front End}
\label{sect:frontend}

\begin{figure*}[t]
    \centering
    \includegraphics[width=1.0\linewidth]{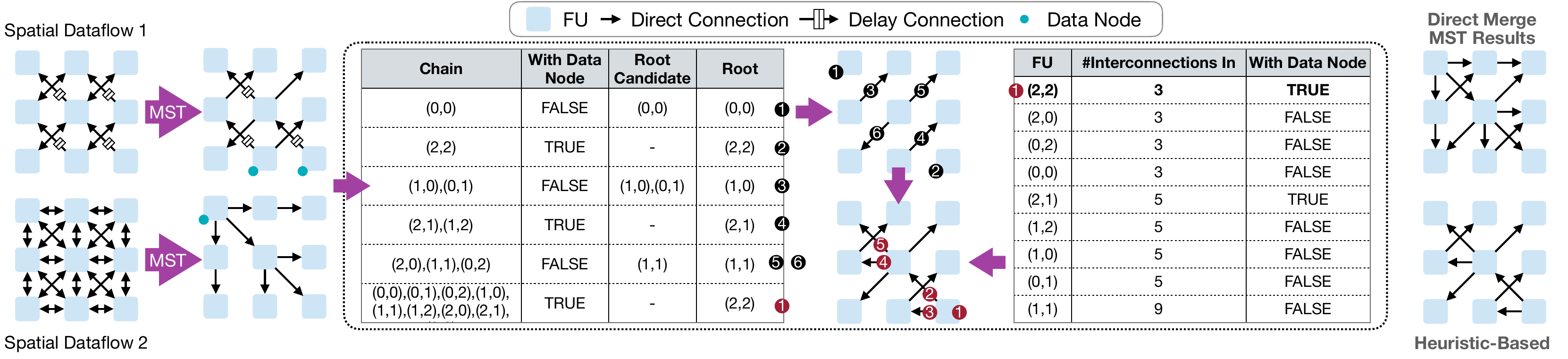}
    \vspace{-20pt}
    \caption{Heuristic-based Direct Interconnections Planning: 1. build interconnections for shorter chain first; 2. select FUs with input delay interconnections as possible root candidates of the chain; 3. select all FUs in the chain as possible root candidates if step 2 fails; 4. select FU with least \#input direct interconnections and data node as final root; 5. extend the ongoing chain with a breadth-first search to connect the longest-built chains.}
    \vspace{-9pt}
    \label{fig:greedy}
\end{figure*}

LEGO front end focuses on establishing FU interconnections to exploit data reuse, building the memory system according to data access patterns, and generating an FU-level architecture description graph as an intermediate representation for back-end processing.

\subsection{Relation-based Interconnection Analysis} 
\label{sect:frontend-interconnection-analysis}
The first step to build potential interconnections between FUs is checking if two FUs access the same data $\vec{d}$.
There are two types of interconnection:
\paragraph{Direct Interconnection} when two FUs always access the same data at the same timestamp, we could directly connect them. From \eqns{eqn:data-mapping} and \ref{eqn:dataflow}, we solve  $\vec{\Delta s}$ that satisfies,
\begin{align}\footnotesize
    \begin{split}
            f_{TS\rightarrow D}(\vec{t}, \vec{s} + \vec{\Delta s}) - f_{TS\rightarrow D}(\vec{t}, \vec{s}) &= \mathbf{0} \\
            \iff \matrixIDmath \matrixSImath \vec{\Delta s} &= \mathbf{0}\\
            constraint: \;\left|\vec{\Delta s}\right|_{\infty}\leq d_{S}, &\Delta t_{bias, \vec{\Delta s}} \ge 0
    \end{split}
    \label{eqn:direct-interconnection}
\end{align}
Each solution $\vec{\Delta s}$ indicates that any two FUs $\vec{s}$ and $\vec{s}+\vec{\Delta s}$ will access the same data at the same timestamp. The constraint $\left|\vec{\Delta s}\right|_{\infty}\leq d_{S}$ limits the spatial distance between these two FUs, and $\Delta t_{bias, \vec{\Delta s}} \ge 0$ makes sure that the data are always shared from past to future (after $\Delta t_{bias, \vec{\Delta s}}$ cycles).

\noindent\textbf{Example:} In \fig{fig:gemm-notation-example}(c), we find a solution $\vec{\Delta s}=(0,1)$ for tensor X. Therefore, FU(0,0) pushes tensor X data to FU(0,1)=(0,0)+(0,1) and FU(1,0) pushes X to FU(1,1). Another solution $\vec{\Delta s}=(0,-1)$ is infeasible as it does not satisfy $\Delta t_{bias} \ge 0$.
Similarly, we could find a solution $\vec{\Delta s}=(1,0)$ for tensor Y.

\paragraph{Delay Interconnection} when two FUs access the same data with a given timestamp gap $\vec{\Delta t}$, we could connect them using a delay FIFO. From \eqns{eqn:data-mapping} and \ref{eqn:dataflow}, we solve the following equation,
\begin{align}\footnotesize
    \begin{split}
        f_{TS\rightarrow D}(\vec{t} + \vec{\Delta t}, \vec{s} + \vec{\Delta s}) - f_{TS\rightarrow D}(\vec{t}, \vec{s}) &= \mathbf{0}\\
        \iff \matrixIDmath \matrixTImath\vec{\Delta t} + \matrixIDmath\matrixSImath \vec{\Delta s} &= \mathbf{0}\\
        constraint: \;\left|\vec{\Delta s}\right|_{\infty}\leq d_{S}, \Delta t_{bias, \vec{\Delta s}} \ge 0
    \end{split}
    \label{eqn:delay-interconnection}
\end{align}
Each solution pair $\vec{\Delta s}$, $\vec{\Delta t}$ indicates that FU $\vec{s}+\vec{\Delta s}$ at timestamp $\vec{t}+\vec{\Delta t}$  will access the same data as FU $\vec{s}$ at timestamp $\vec{t}$. The FIFO depth can be derived from $\vec{\Delta t}$ following \eqn{eqn:time-scalar}.

\noindent\textbf{Example:} In \fig{fig:gemm-notation-example}(c), we get a solution pair $\vec{\Delta s} = (0,-1)$ and $\vec{t}=(0,\cdots,0,1)$ for tensor X. Therefore, FU(0,1) pushes tensor X data to FU(0,0)=FU(0,1)+(0,-1) and FU(1,1) pushes X to FU(1,0) with one cycle delay. We also get a solution pair $\vec{\Delta s} = (-1,0)$ and $\vec{t}=(0,\cdots,1,0)$ for tensor X. Therefore, FU(1,0) also pushes tensor X data to FU(0,0)=FU(1,0)+(-1,0). 
Since both solutions are delay connections, the tensor X data received from FU(0,1) will be valid if and only if the timestamp of FU(0,0) is no smaller than $(0,\cdots,0,1)$, and data received from FU(1,0) will be valid if and only if the timestamp of FU(0,0) is no smaller than $(0,\cdots,1,0)$.

\paragraph{Differences to TensorLib}
Previous work TensorLib{~\cite{jia2021tensorlib}} establishes all interconnections by solving the data reuse hyperplane. However, TensorLib only considers hyperplane up to rank = 2 (\ie, 2D array) and simplifies designs to only 7 cases by allowing only one delay interconnection set. In the CONV2D example of Figure~{\ref{fig:conv-notation-example}}, Tensorlib can only generate one possible interconnection set for tensor X while ignoring the other. Furthermore, the TensorLib representation does not take the loop tiling into consideration: the parfor-loop and for-loop nests may partition the same computation iteration and thus change the FU interconnection accordingly. In contrast, LEGO neither limits the number of spatial dimensions, nor limits the number of delay interconnection sets.

\subsection{Minimum-Spanning Interconnections Generation}

As depicted in the leftmost of \fig{fig:greedy}, FU interconnections found in the previous step may be excessive.
Therefore, to guarantee that each FU will have only one valid data source for each tensor operand at any timestamp,
we will search the minimum spanning trees
to get the minimum set of necessary connections.
The cost of each edge is the depth of delay FIFO, so the MST algorithm could be used to minimize the overall cost introduced by delay connections.
Since the entire graph is directed, we use Tarjan's Chu-Liu algorithm~\cite{tarjan1977finding}. 
The root FU of every tree will be labeled with a data node indicating it requires fetching/committing the data from/to memory.

\subsection{Heuristic-based Direct Interconnections Planning}

A program in real life, such as deep learning neural networks, may have multiple tensor kernels on the same hardware and these kernels may have different spatial dataflows.
If the expected hardware supports multiple spatial dataflows, we have to combine interconnections under different dataflows into one graph. However, directly merging all the minimum-spanning interconnections obtained from the previous step 
will not yield the optimal solution, especially for direct interconnections. Therefore, we propose a breadth-first search (BFS) based heuristic algorithm to re-establish all direct interconnections as illustrated in \fig{fig:greedy}.

We partition FUs in the graph of each spatial dataflow (\matrixSI): all FUs in the same subgraph can be connected to each other with direct interconnections. We refer to such subgraph as a \textit{chain}. The FU providing the shared data is referred to as the \textit{root} of the chain. The key of the heuristic rules is to 1) try reusing short broadcast chains to form a longer chain to reduce the number of muxes and thus the complexity of FU interconnections, and 2) try reusing FUs with the data node as the root in the chain to reduce the number of data nodes and thus the complexity of data distribution switches.

First, we record the FUs receiving data with delayed interconnections as the root candidates.
In the spatial dataflow 1 of \fig{fig:greedy}, FU(2,0), FU(1,1), and FU(0,2) are connected with direct connections, and thus form a chain. Since only FU(1,1) can pull data from the delay connection, FU(1,1) becomes the only root candidate of this chain.

Then, we build direct interconnections for each chain one by one from longer to shorter. The root of the ongoing chain is selected from its root candidates: the one having the least number of possible input direct interconnections. Starting from the root, we extend the ongoing chain with a breadth-first search to connect the longest-built chains. For example, when working on the longest chain in \fig{fig:greedy}, all FUs are the root candidates. Among them, only FU(2,2) has 3 input direct interconnections and is labeled with a data node in the previous shorter chains (the chain with \blackcircle{2}). Therefore, FU(2,2) will be selected as the root of this chain. Among all neighbors of root FU(2,2), both FU(1,1) and FU(2,1) are the roots of other chains. Since the chain with FU(1,1) as the root is longer, FU(1,1) will be first connected (\redcircle{2} in the figure). We will repeat this process until all FUs are connected.

After determining the direct interconnections in the previous step, we will add delay interconnections between the root FUs of direct interconnection chains accordingly. 
\begin{figure}[t]
    \centering
    \includegraphics[width=\linewidth]{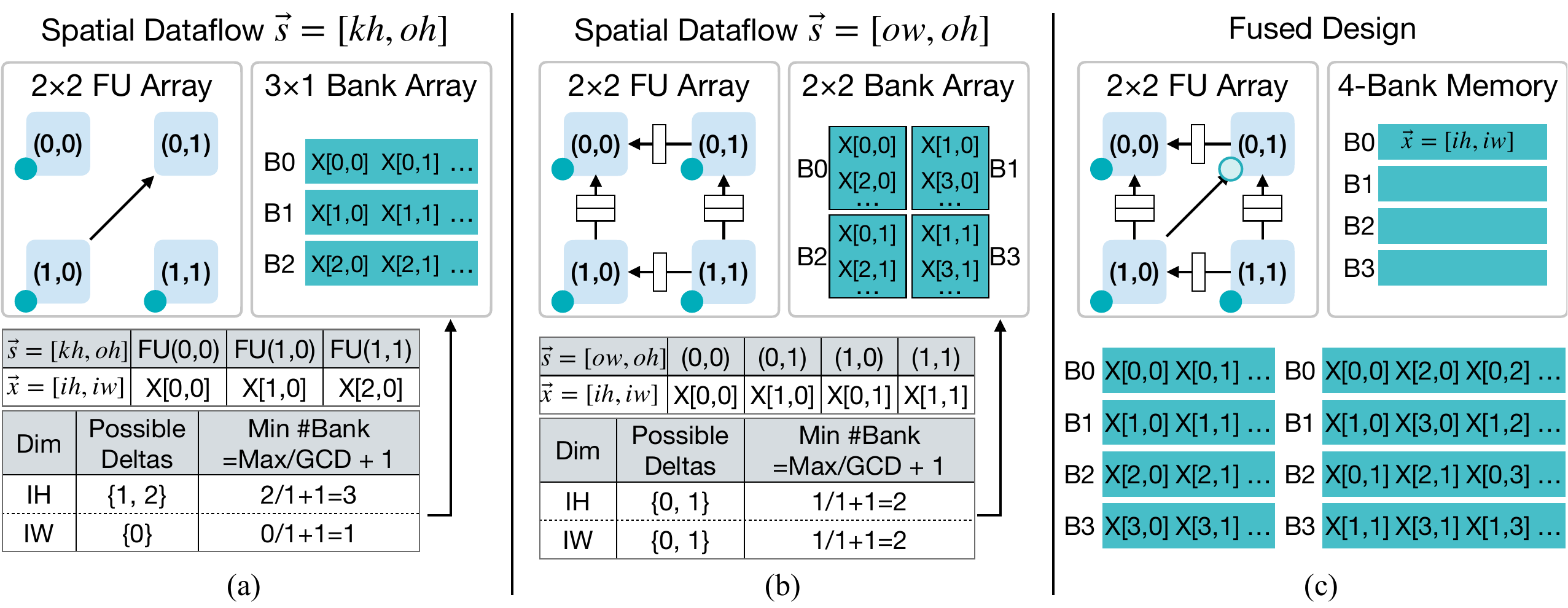}
    \vspace{-18pt}
    \caption{Data access patterns of FUs determine the tensor data layout in the on-chip memories. We examine the tensor indexes of all data nodes at $\vec{t}=0$. As long as the number of banks in each dimension is larger than the maximum possible delta in tensor indexes, no bank conflicts will occur. 
    }
    \label{fig:memory-distribution}
    \vspace{-12pt}
\end{figure}

\subsection{Relation-based Memory Analysis}
\label{sect:frontend-memory-allocation}

As discussed in Section{~\ref{sect:overview}}, LEGO decouples the L1 memory system and FU array by using the data distribution switches to direct data between buffers and FUs. 
The buffers are divided into banks, forming a $n_D$-D memory array for a $n_D$-d tensor. The bank index $\vec{b}$ for the tensor data index $\vec{d}$ can be calculated by an element-wise modulo $b_i = d_i\mod B_i$, where $B_i$ is the number of banks in dimension $i$. 

Since multiple FUs may require access to memory simultaneously, we have to avoid bank conflicts to prevent stalls, \ie for any two FUs labeled with data node, $s_1$ and $s_2$
\begin{equation}
\vec{d_1} - \vec{d_2} = f_{TS\rightarrow D}(\vec{t}, \vec{s_1}) - f_{TS\rightarrow D}(\vec{t}, \vec{s_2}) \not\equiv 0  
\  (mod \  \vec{B})
\end{equation}

Without loss of generality, we could set $\vec{t}=0$, and examine the delta of tensor index. The interconnection analysis in \ref{sect:frontend-interconnection-analysis} ensures $\vec{d_1} \neq \vec{d_2}$. As long as 
\begin{equation}
B_i > \max_{\vec{s_1}, \vec{s_2}} (\vec{d_1}_i - \vec{d_2}_i) = \max_{\Delta\vec{s}}(\Delta d_i)
\end{equation}
no bank conflict will occur. 
This can be achieved by enumerating all possible $s_1$ and $s_2$ pairs in the FU array and setting $B_i$ to $\texttt{Max}(\{|\Delta d_i|\}) + 1$.

Additionally, if we find these deltas have a common divisor $g_i = \texttt{GCD}(\{|\Delta d_i|\})$, we can further reduce the number of banks by letting $B_i$ to be $\texttt{Max}(\{|\Delta d_i|\})/\texttt{GCD}(\{|\Delta d_i|\}) + 1$, and thus the bank index is calculated as $b_i = d_i / g_i \mod B_i$. 

After allocating the tensor data on on-chip memory, data distribution switches will be built between FU data nodes and on-chip memory to ensure sufficient on-chip bandwidth for providing the tensor data during execution. 

\noindent\textbf{Example}. In \fig{fig:memory-distribution}(a), there are only 3 FUs labeled with data nodes for tensor X. At $\vec{t}=0$, the accessed tensor data are X[0,0], X[1,0], and X[2,0]. The possible deltas of tensor X index are [2,0] and [1,0], \ie, $\{\Delta d_{\texttt{IH}}\} = \{1, 2\}$ and $\{\Delta d_{\texttt{IW}}\} = \{0\}$. Thus we could set $B_{\texttt{IH}}$ to at least 3, and $B_{\texttt{IW}}$ to at least 1, resulting in a memory array of 3 banks. In \fig{fig:memory-distribution}(b), the possible deltas are [1,0], [0,1], and [1,1], so we set both $B_{\texttt{IH}}$ and $B_{\texttt{IW}}$ to at least 2, resulting in a 2×2 memory array. When fusing (a) and (b), we can use a 4-bank memory system which is viewed as 4×1 banks for (a) and 2×2 banks for (b).

\section{LEGO Back End}
\label{sect:backend}

The architecture description graph (\adg) generated by the LEGO front end provides the hardware connection information at a relatively high level, of which the basic components are FUs, data nodes, and connections. However, they are not the elementary components - an FU may perform different operations in different dataflows; a connection needs to be implemented in FIFOs, wires, broadcast, or reduction tree; multiple connections to a single FU need to be merged by MUXes or reduction logic.
\adg ignores the data types and bitwidths of signals and uses an ideal hardware model assuming everything is a combinational logic.
This simplifies the front-end analysis but is not realistic.

Prior works like AutoSA~\cite{wang2021autosa} and Tensorlib~\cite{jia2021tensorlib} tend to use manually designed FU templates to convert similar higher-level abstractions into HLS/RTL codes. With MST-based interconnection generation and fusion of multiple dataflows, we found that the structure of FUs and their connections has become too complex to design and optimize by hand. Relying on synthesis/HLS tools is also a suboptimal choice since they cannot change the semantics of the logic and may prevent some optimizations. Therefore, we propose LEGO back end that systemically generates RTL from \adg.

\begin{figure}[t]
    \centering
    \includegraphics[width=0.675\linewidth]{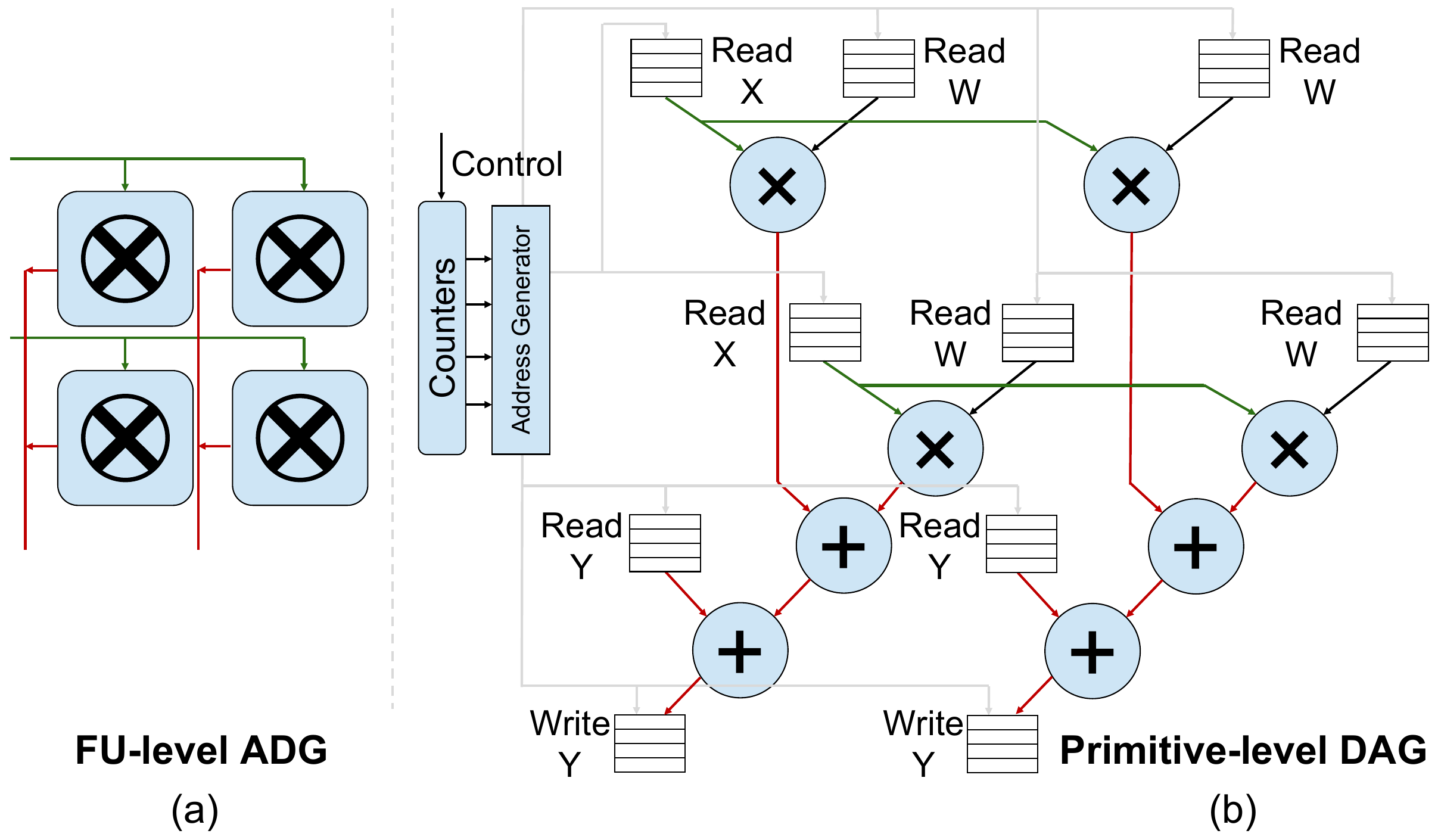}
    \vspace{-9pt}
    \caption{An example of an architecture description graph (ADG) (a) and its corresponding detailed architecture graph (DAG) (b). ADG describes the hardware at the FU level and defines the connections in between, while DAG opens the black boxes of FUs and describes the hardware using basic primitives.
    The boundaries of FUs are also eliminated in DAG.
    }
    \vspace{-12pt}
    \label{fig:backend-dag}
\end{figure}
\begin{figure*}[t]
    \centering
    \includegraphics[width=0.95\linewidth]{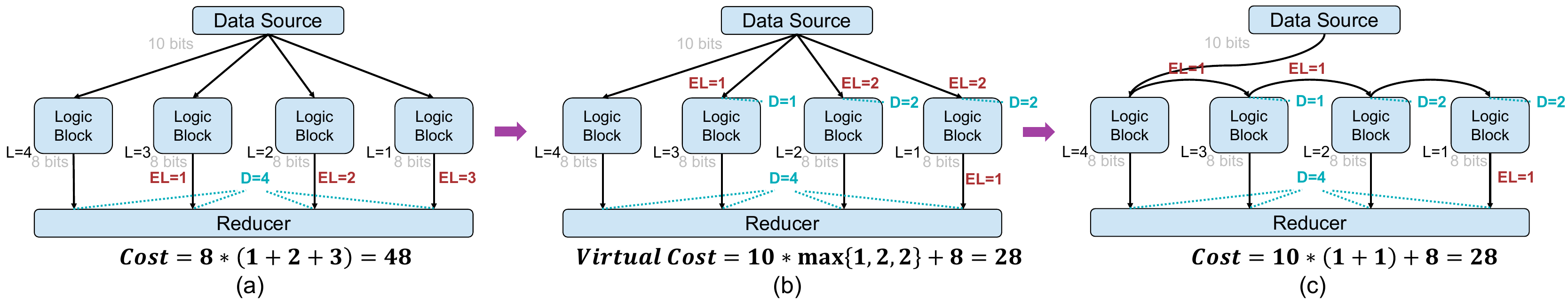}
    \vspace{-10pt}
    \caption{Pin rewiring. The pipeline registers (denoted by EL) inserted during delay matching can be suboptimal when there are broadcast pins, and the broadcast may not be directly transformed into a forwarding chain in the case of (a). In (b), we modify the cost function of the linear program to a virtual one that only counts the maximum cost for each broadcast pin. This encourages it to put the pipeline register directly after the broadcast, which allows us to rewire the broadcast connection in (c) using MST.}
    \label{fig:rewire}
    \vspace{-9pt}
\end{figure*}

LEGO back end introduces a lower-level representation - a detailed architecture graph (\dag).
\fig{fig:backend-dag} shows an example of \dag. 
Unlike \adg, which adopts FUs as its nodes, \dag selects more low-level hardware primitives as its node. 
The edges between nodes describe the properties including data range, bit-width, and data delay in cycles.

A remarkable node is the address generation module, which maps the current timestamp to the data address for each tensor. The timestamp generation is achieved by a series of counters with carry logic, and the mapping is achieved by matrix multiplication. Thanks to the affine-transformation-based representation, we could always write the address generation logic as matrix multiplication with bias, and when dataflow changes, we only need to change the values in the matrix without changing the hardware structure, which simplifies the hardware design and provides more flexibility.

LEGO back end performs a translation pass (codegen) on \adg to build the \dag and then uses multiple transformation passes on the \dag to optimize and elaborate the hardware, the goal of which is to reduce register usage. This is based on the observation that a majority of resource consumption comes from the data path: in the FU array, registers consume \>40\% of the total area and \>60\% of the total power.

\subsection{Delay Matching on \dag}

\adg ignores the internal latencies of hardware components, when taking that into consideration, we need to add extra pipeline registers to make sure every component can get the data at the correct time. This requires all paths to the input pins of a component to have the same delay. The inserted pipeline register directly contributes to the hardware resource consumption, so we formulate this as an optimization problem: 
let $D_v$ denotes the delay of the output of node $v$, $L_v$ denotes the internal latency of $v$, and $EL_{u,v}$ denotes the number of inserted pipeline registers on the edge of $(u,v)$, we have $D_v=D_u+EL_{u,v}+L_v$ for all $(u,v)$ in DAG. The constraint is that $EL_{u,v}$ cannot be negative, so
\begin{equation}
EL_{u,v}=D_v-D_u-L_v\ge0
\end{equation}
and the optimization target is the total number of registers inserted into DAG, which is
\begin{equation}\footnotesize
\min_{D} \sum_{(u,v)\in E} EL_{u,v}*W_{u,v}\\
= \min_{D} \sum_{(u,v)\in E} (D_v-D_u-L_v)*W_{u,v}
\end{equation}
where $W_{u,v}$ is the bit-width of edge $(u,v)$. This linear programming problem can be solved with any LP solver.

\subsection{Broadcast Pin Rewiring on \dag}

On \dag, multiple signals may share the same data source,
and finally be connected to a reduction logic after several computation logics. Converting a broadcast connection to a forward connection can save the number of pipeline registers.

However, in practice, the situation can be different as the linear programming solver may not explicitly incorporate the latency on the broadcasted signal, as shown in \fig{fig:rewire} (a).
To solve this problem, we propose a three-stage heuristic algorithm for finding good rewiring.

In the first stage, the cost function of the linear program is modified to give preference to the broadcasted signal when adding latency.
We use an optimistic estimation of the cost of a broadcasted signal, which is set to the maximum latency between the source and all its destinations, since without spatial constraint we could always convert the broadcast into a chain.
\fig{fig:rewire} (b) shows the process of cost adjustment. 

The second stage uses a Minimum Spanning Tree (MST) solver for each broadcast source $s$ to perform the rewiring. First, an edge is created from the original source to each destination $u$ to represent the direct connection, of which the cost is the original latency $E_{s, u}$. Then edges are added between spatially adjacent destinations $u$, $v$ to represent the forwarding, of which the cost is the difference of the delay $|E_{s,u}-E_{s,v}|$. \fig{fig:rewire} (c) shows a possible rewiring solution. 

In the third stage, the linear program is re-run on the rewired \dag. It is optional but can ensure a correct and optimal solution after rewiring by redistributing the extra latencies.

\subsection{Reduction Tree Extraction and Pin Reusing}
\label{sect:pin-reuse}
\begin{figure}[t]
    \centering
    \includegraphics[width=0.85\linewidth]{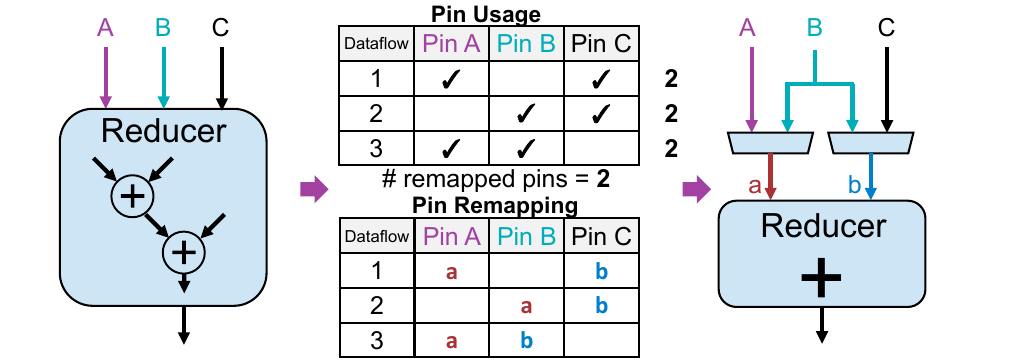}
    \caption{Pin reusing runs after reduction tree extraction. A liveness analysis for each dataflow summarizes a table listing used pins in each case. The number of inputs after remapping will be the maximum number of the used pin in all situations. A pin mapping can be established by a 0-1 integer program, which maps original pins (A, B, C) to the ports (a, b) of reducer. }
    \label{fig:reuse-pin}
    \vspace{-9pt}
\end{figure}
Another transformation pass is the extraction of reduction logic. On \adg, the reduction logic is represented by a long adder chain.
The long sequential logic will result in more registers inserted by the delay matching process. This pass identifies directly connected adders 
and converts them into a single reduction unit which will later become a balanced tree of reduction, which greatly reduces the levels of logic.

The extraction of reduction logic allows us to optimize it further by reusing the pins. In scenarios with multiple dataflow designs, it is possible that not all of the input pins of a reducer are used simultaneously. The input pins can be remapped to minimize the cost of the reducer. A smaller reducer also means fewer logic levels and fewer registers in the datapath.

\begin{figure*}[t]
    \centering
    \includegraphics[page=1, width=0.485\linewidth]{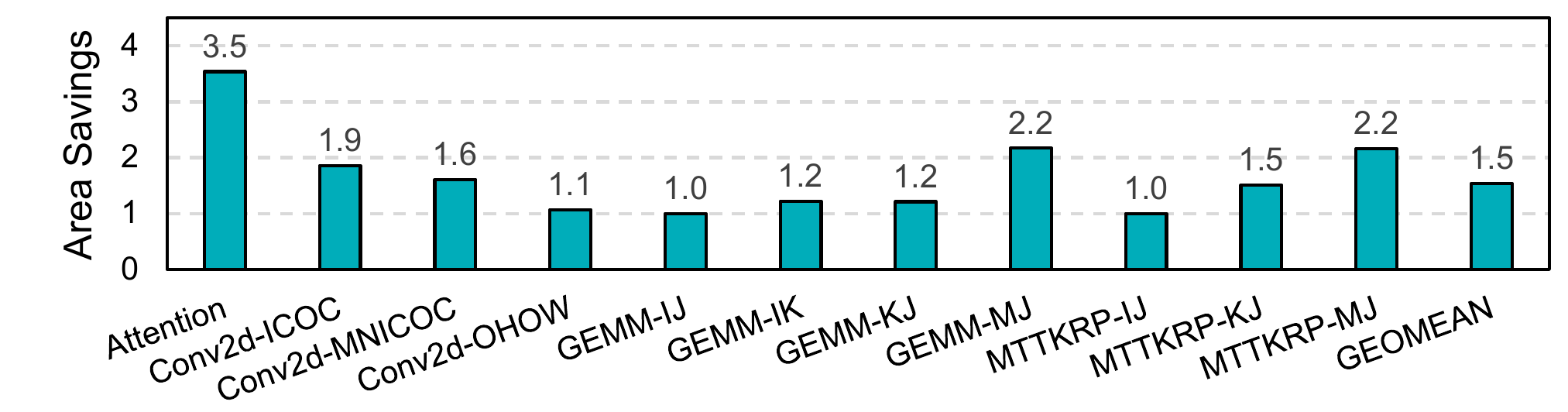}
    \includegraphics[page=2, width=0.485\linewidth]{figure/6-evaluation/evaluation.pdf}
    \vspace{-12pt}
    \caption{The area and energy savings of LEGO optimizations on different tensor kernels with various execution dataflows, 
    named as \textit{Operation-Dataflow}. \textit{M} and \textit{N} represent dynamically switchable dataflow. The baseline (delay matching only, which is mandatory for the timing requirement) is compared with the optimized architecture (with pin reusing, reduction tree optimizations, and power gating).
    }
    \label{fig:savings-kernel}
\end{figure*}
\begin{figure*}
\centering
\includegraphics[page=3, width=0.485\linewidth]{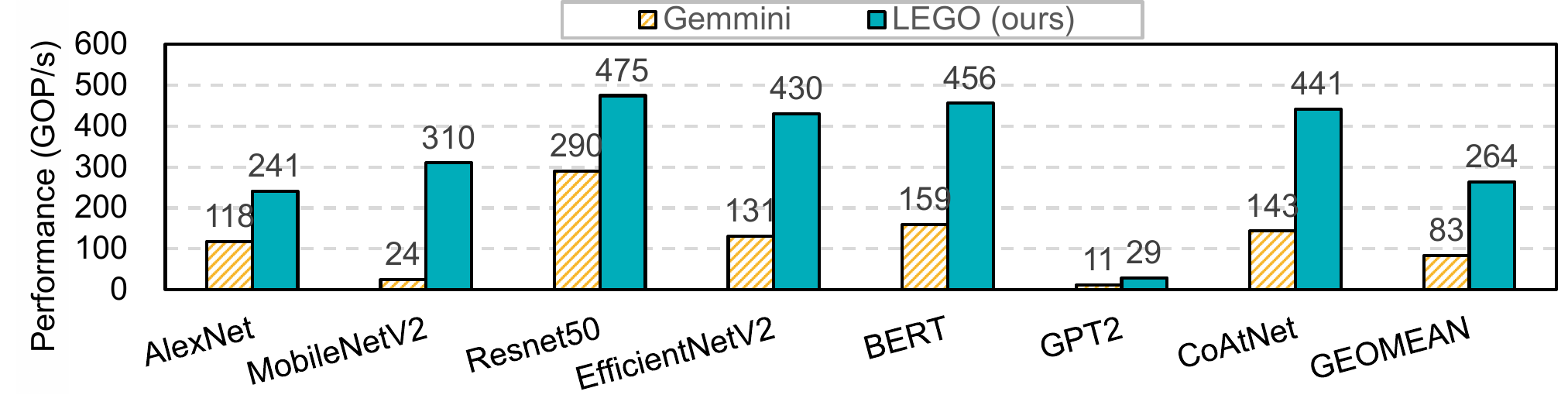}
\includegraphics[page=4, width=0.485\linewidth]{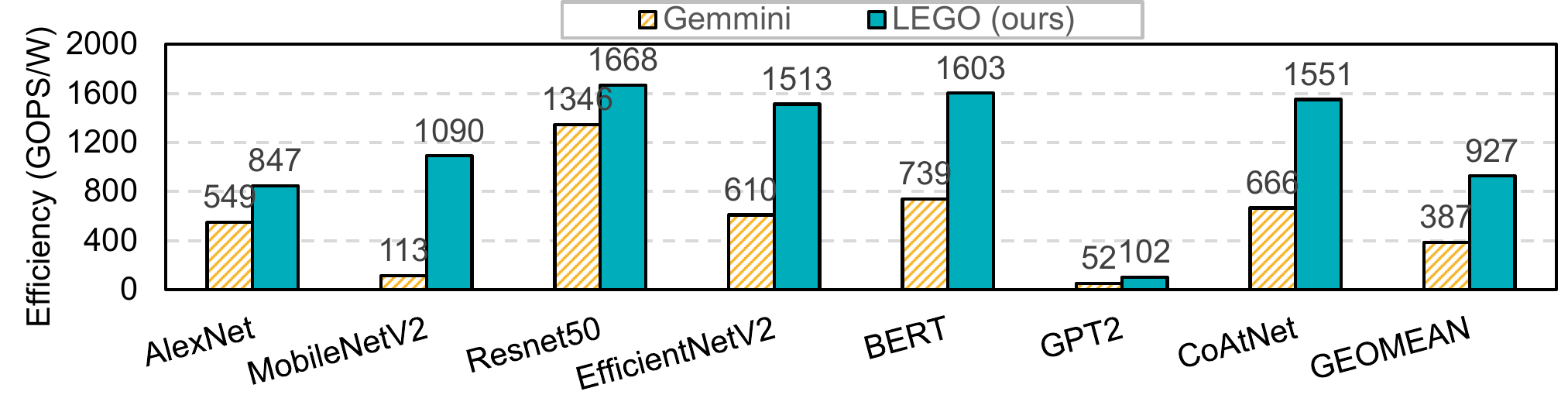}
\vspace{-12pt}
\caption{End-to-end performance and energy efficiency comparison (including the post-processing unit) between Gemmini and LEGO. LEGO achieved an average of {3.2×} speedup and {2.4×} energy savings over Gemmini. Both Gemmini and LEGO are bounded by memory bandwidth on GPT2. LEGO performs much better on MobileNetV2 due to its efficient support of dynamically switching spatial dataflows.}
\vspace{-9pt}
\label{fig:savings-end-to-end}
\end{figure*}

The first step is to identify which input pins are used in each dataflow configuration, achieved through a liveness analysis. 
By traversing the \dag, we can determine whether a pin contributes to any true dependency in the final output (output buffer). The information is recorded in a pin usage table to derive the number of remapped pins, as shown in \fig{fig:reuse-pin}.

The next step is to build a pin remapping. This is done by a 0-1 integer program. We use 0-1 variables $C(i,j,k)$ to control if the original pin $i$ should be mapped to physical pin $j$ in dataflow $k$, and set $A(k)$ to represent the active pins in dataflow $k$. 
We need to make sure each active pin must connect to one physical pin: $\sum_{j} C(i,j,k) = 1$ (if $i \in A(k)$), and each physical pin only connects to at most one input: $\sum_{i} C(i,j,k) \leq 1$. The optimization objective is to minimize the connections $\sum_{i,j,k} C(i,j,k)$. 

The \dag will be built based on the pin mapping. If multiple inputs are mapped to one reducer pin, a MUX will be used. Since a MUX is much more lightweight than an adder on ASIC, the pin reusing could effectively reduce area and power.

\subsection{Other Transformation and Elaboration Passes on \dag}

The design of \dag enables many useful elaboration and optimization passes. For example, a power gating pass can reduce the power consumption of unused connections between FUs by adding clock-enabled signals before each delay block. Another important pass is the bit-width inference, which calculates the value range of each signal and determines the bitwidth information on \dag.

\section{Evaluation}
\label{sect:evaluation}

\subsection{Evaluation Methodology}

We implemented the LEGO framework in C++ with HiGHS~\cite{huangfu2018parallelizing} as the linear programming solver in the back end. All of the \dag primitives and NoC components are built in SpinalHDL to generate the synthesizable Verilog code from the optimized \dag.
We synthesized the generated hardware using Synopsys Design Compiler with TSMC 28nm library and used CACTI~\cite{muralimanohar2009cacti} to model SRAM. 
We also built a performance simulator for the FU array and NoC in the \frontend to fast predict the latency of computation and memory movement, which are verified with the RTL simulation.

We evaluated LEGO on four crucial tensor kernels with different dataflows: GEMM, Conv2d, Attention, and Matricized tensor times Khatri-Rao product (MTTKRP). In addition to GEMM, Conv2d is the building block in CNNs, Attention is a pivotal operation in transformer-based foundation models, and MTTKRP is the bottleneck in tensor factorization such as alternating least squares (ALS) in recommendation systems.

We also evaluated LEGO with neural networks, including classical convolution neural networks AlexNet~\cite{krizhevsky2017imagenet}, MobileNetV2~\cite{sandler2018mobilenetv2}, Resnet50~\cite{he2016deep}, EfficientNetV2~\cite{tan2021efficientnetv2}, transformer-based models BERT~\cite{devlin2018bert}, GPT-2~\cite{radford2019language}, CoAtNet~\cite{dai2021coatnet}, and modern generative AI models including DDPM~\cite{ho2020denoising}, Stable Diffusion~\cite{rombach2022high}, and LLaMA-7B~\cite{touvron2023llama}. All non-tensor functions are run on post-processing units.
The BERT sentence length is 16. The prompt length of GPT-2 and LLaMA-7B is 1000 and the latency of generating one token is measured. The image size is 384×384×3 on EfficientNetV2, and 224×224×3 on other vision models.

In tensor kernel evaluation, LEGO generated different designs for different dataflows. All designs generated by LEGO are denoted by their spatial parallelism dimensions. \textit{M} and \textit{N} represent switchable spatial dimensions. For example, ``GEMM-MJ'' indicates the generated hardware supported both I-J parallelism and K-J parallelism for the GEMM operation. In the end-to-end evaluation, LEGO generated only \textit{one} single hardware (``LEGO-MNICOC'') across \textit{all} NN models. We implemented a simple mapping search tool that identifies the best mapping (\ie, dataflow and tiling) for every neural network layer based on the simulated \#cycles and energy from LEGO front end.

For end-to-end performance comparison, we adopted the state-of-the-art open-source NN accelerator generator, Gemmini~\cite{genc2021gemmini}, as our major baseline, 
since it provides a complete end-to-end workflow and delivers comparable performance to our work.
For a fair comparison, we configured LEGO to use hardware resources similar to Gemmini: 256 MACs, 256 KB on-chip buffer, and a 128-bit memory bus with 16GB/s bandwidth. We also minimize the system overhead in Gemmini by only counting the \#cycles of the tensor kernel itself. We also compared against other state-of-the-art works, including HLS-based SODA optimizer~\mbox{\cite{agostini2022soda}} and FPGA-targeted AutoSA~\mbox{\cite{wang2021autosa}}, and the results are summarized in Table~{\ref{tab:comparison-related}}. When comparing with SODA, LEGO generated a tiny FU array with 16 FUs and synthesized with the same FreePDK 45nm technology as SODA. When comparing with AutoSA, both AutoSA and LEGO generated varying designs with an 8\mbox{×}8 FU array for multiple tensor kernels and synthesized on Xilinx Ultrascale+ U280 FPGA.

\subsection{Experimental Results}
\label{sect:evaluation-results}

\paragraph{Tensor Kernel Performance} 
As shown in \fig{fig:savings-kernel}, LEGO supports various tensor kernels with different numbers of input tensors, different computation kernels, and different spatial dataflows. LEGO is also able to fuse different spatial dataflows in one design
(e.g., score-stationary Attention{~\cite{lu2021sanger}}, Conv2d-MNICOC, MTTKRP-MJ).
On average, LEGO achieves {1.5×} area savings and {1.4×} energy savings. The improvement is especially remarkable for dynamic dataflows since connection topology in these designs is more complex and provides more opportunities for optimization.

\paragraph{End-to-End Performance}
\fig{fig:savings-end-to-end} compares the end-to-end performance and energy efficiency improvement running neural network models with Gemmini and LEGO, with all non-tensor functions, including activation functions and normalization, running on PPUs.
LEGO achieves nearly theoretical maximum performance on Resnet50, EfficientNetV2, BERT, and CoAtNet models. Both Gemmini and LEGO do not perform well on GPT-2 since generative models like GPT-2 are typically bounded by memory bandwidth. On average, LEGO achieves 3.2× speedup and 2.4× energy saving over Gemmini. The speedup mainly comes from 2 parts: (i) LEGO front end integrates a fast and accurate performance modeling tool, which helps guide the scheduler to search the optimal mapping policy for the hardware; (ii) the support of flexible FU connection generation enables designs with dynamic spatial dataflow switching, which extends the search space for the scheduler. For example, the speedup is significant on depthwise convolutional layers since LEGO could switch to OH-OW-IC-OC dataflow on these layers.

\begin{figure}
\begin{minipage}{\linewidth}
\end{minipage}
\begin{minipage}{\linewidth}
    \centering
    \includegraphics[page=5,width=\linewidth]{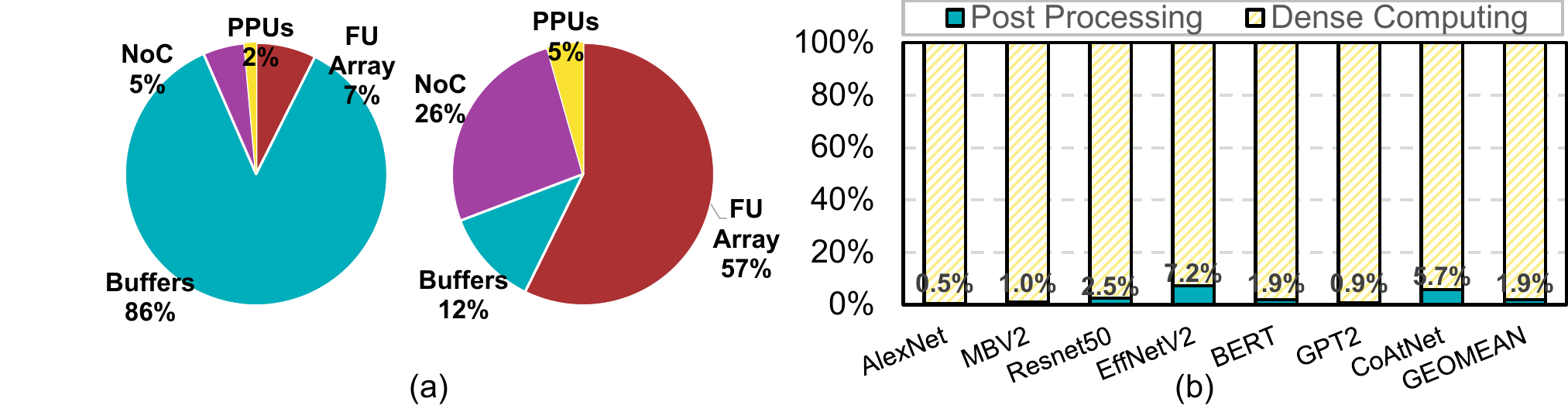}
    \vspace{-16pt}
    \caption{(a) Area (left, 1.76mm$^2$ in total) and on-chip power (right, 285mW in total) breakdown of LEGO-MNICOC.
    Post-processing units (PPUs) take up to 2\% area overhead and 5\% power consumption. (b) The end-to-end latency breakdown. The post-processing unit processes the results before writing back to the DRAM, which avoids the cost of CPU communication.
    }
    \label{fig:hardware-breakdown}
    \vspace{-9pt}
\end{minipage}
\end{figure}
\begin{table}[t]
\centering
\caption{
Large generative models on LEGO-ICOC-1K with 1024 FUs, 576 KB buffer, 32 post-processing units, and 32GB/s bandwidth.
The on-chip area and power are 3.95 $mm^2$ and 601 mW.
}


\scriptsize
\scalebox{1.0}{
\begin{tabular}{@{}c|c|c|c|c@{}}
\toprule
\multirow{2}{*}{\textbf{Model}} & \multirow{2}{*}{\textbf{DDPM}} & \textbf{Stable}    & \textbf{LLaMA-7B} & \textbf{LLaMA-7B} \\
                                &                                & \textbf{Diffusion} & bs=1              & bs=32             \\ \midrule
Utilization                     & 92.9\%                         & 80.2\%             & 3.1\%             & 42.9\%            \\ \midrule
Perf. (GOP/s)                   & 1903                           & 1642               & 63                & 878               \\ \midrule
Energy Eff. (GOP/s/W)            & 3165                           & 2731               & 105               & 1461              \\ \bottomrule
\end{tabular}
}
\vspace{-9pt}
\label{tab:large-models}
\end{table}

For generative AI models, a larger and simpler architecture is adopted with the same computation resource (1024 \#FUs). As shown in \tab{tab:large-models}, LEGO achieves $>$80\% utilization on DDPM and Stable Diffusion. LLaMA-7B model has very low operational intensity, especially in attention layers, of which the performance is heavily bounded by the DRAM bandwidth.

\paragraph{Comparison to Human Design with Handwritten RTL}
\tab{tab:compare-eyeriss-nvdla} compares the quality of the generated hardware by LEGO against expert-designed accelerators, including Eyeriss~\cite{chen2016eyeriss} and NVDLA~\cite{zhou2018research}, using the same dataflows and settings. The automatically generated implementation achieves comparable area and power. Additionally, LEGO achieves lower power consumption than Eyeriss, thanks to the data read from buffers being shared among FUs via the FU interconnections, reducing power in scratchpads.

\begin{table}[t]
\centering
\caption{Comparison b/w handwritten and LEGO-generated designs.
}
\scriptsize
\scalebox{1.0}{
\begin{tabular}{@{}l|c|c|c|c@{}}
\toprule
\textbf{Architecture}   & \textbf{Eyeriss}          & \textbf{LEGO-KHOH} & \textbf{NVDLA}            & \textbf{LEGO-ICOC}       \\ \midrule
Dataflow & \multicolumn{2}{c|}{KH-OH Parallel} & \multicolumn{2}{c}{IC-OC Parallel} \\ \midrule
\#FUs          & \multicolumn{2}{c|}{168}   & \multicolumn{2}{c}{256} \\ \midrule
Frequency      & \multicolumn{2}{c|}{200 MHz} & \multicolumn{2}{c}{1 GHz}    \\ \midrule
Technology     & 60nm  & 65nm  & 28nm  & 28nm  \\ \midrule
Area (mm$^2$)  & 9.6   & 7.4   & 1.7   & 1.5    \\ \midrule
Power (mW)     & 278   & 112   & 300*  & 209     \\ \bottomrule
\multicolumn{5}{l}{* projected from 16nm~\cite{wu201316nm}}
\end{tabular}
}
\vspace{-9pt}
\label{tab:compare-eyeriss-nvdla}
\end{table}
\begin{table}[t]
\centering
\caption{Runtime cost and performance when scaling up design.
}

\scriptsize
\begin{tabular}{@{}c|c|c|c|c|c@{}}
\toprule
\#FUs                & 64   & 256   & 1024  & 4096  & 16,384  \\ \midrule
FU Array (max 1024 FUs)           & 8×8  & 16×16 & 32×32 & 32×32 & 32×32 \\ \midrule
L2 NoC Size         & 1    & 1     & 1     & 2×3   & 4×5     \\ \midrule
Generation Time (s) & 13.1 & 28.7  & 111.2 & 120.3 & 134.3   \\ \midrule
Area ($mm^2$)       & 0.02 & 0.06  & 0.24  & 1.05  & 4.21    \\ \midrule
Power (mW)          & 29   & 106   & 422   & 1748  & 6987    \\ \midrule
Energy Eff. (GOP/s/W)      & 4404 & 4816  & 4853  & 4688  & 4690    \\ \bottomrule
\end{tabular}
\vspace{-9pt}
\label{tab:scaleup}
\end{table}
\begin{table}[t]
\centering
\caption{Efficacy of fusing multiple dataflow in a single design.
}

\scalebox{0.75}{
\begin{tabular}{@{}c|c|c|c|c@{}}
\toprule
\multirow{2}{*}{\textbf{Architecture}} & \textbf{LEGO-}    & \textbf{LEGO-}    & \textbf{LEGO-}  & \textbf{LEGO-}  \\
                                       & \textbf{ICOCICOC} & \textbf{OHOWICOC} & \textbf{MNICOC} & \textbf{MNICOC} \\ \midrule
\#Spatial Dataflow                               & Single            & Single            & Both        & Both        \\ \midrule
FU Connections Generation              & Regular           & Regular           & Simply Merged   & Optimized       \\ \midrule
Power (mW)                             & 123               & 155               & 196             & 163             \\ \midrule
MBV2 Perf. (GOP/s)                          & 213               & 293               & 313             & 313             \\ \midrule
MBV2 Eff. (GOP/s/W)                         & 1732              & 1890              & 1597            & 1920            \\ \midrule
Resnet50 Perf. (GOP/s)                          & 409               & 422               & 487             & 487             \\ \midrule
Resnet50 Eff. (GOP/s/W)                         & 3325              & 2723              & 2485            & 2988            \\ \bottomrule
\end{tabular}
}
\label{tab:ablation-frontend}
\vspace{-9pt}
\end{table}

\begin{figure}[t]
\begin{minipage}{\linewidth}
    \centering
    \includegraphics[page=1,width=\linewidth]{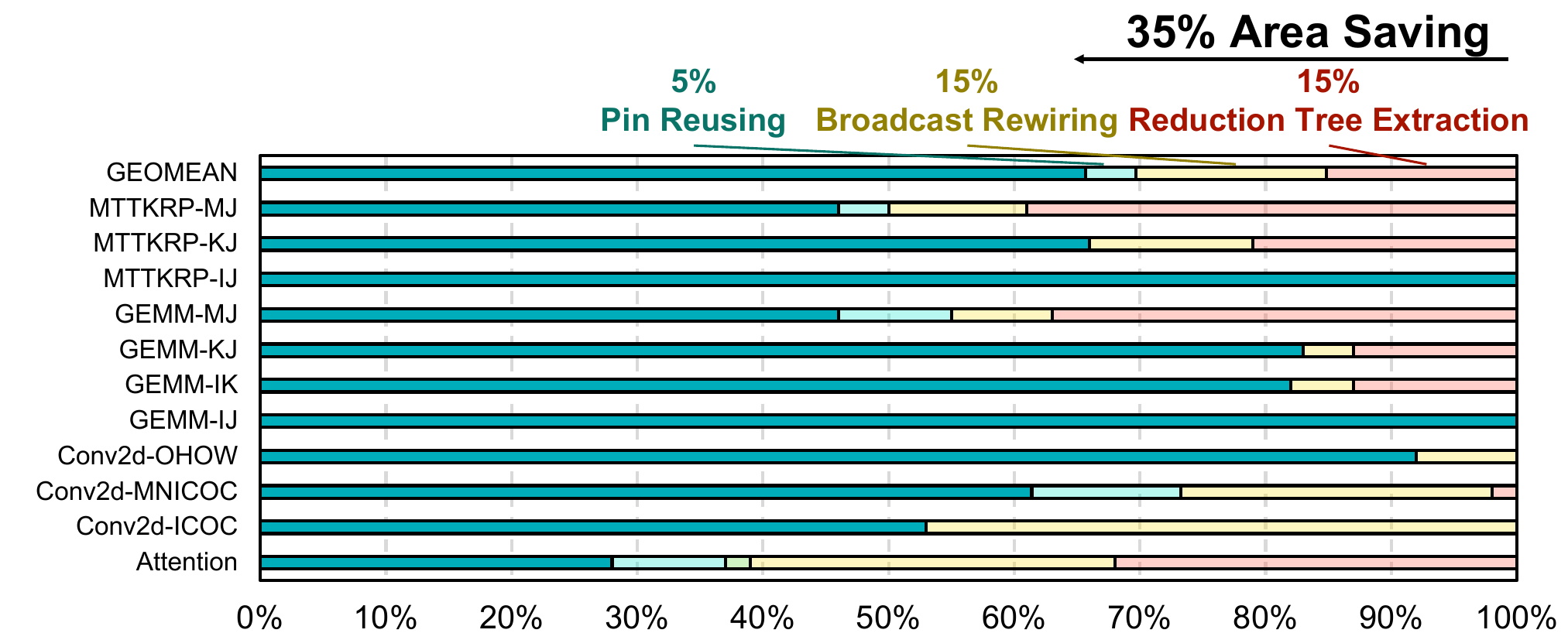}
    \vspace{-16pt}
    \caption{Performance breakdown for LEGO backend optimizations on single tensor kernels. LEGO backend can optimize the area by 35\% on average with significant improvement on designs with switchable dataflows (\eg, MTTKRP-MJ, Conv2d-MNICOC, Attention).}
    \label{fig:optim-breakdown-area}
\end{minipage}
\begin{minipage}{\linewidth}
    \centering
    \includegraphics[page=2,width=\linewidth]{figure/6-evaluation/perf-breakdown.pdf}
    \vspace{-16pt}
    \caption{Power breakdown for LEGO backend optimizations. LEGO backend optimizes the power by 28\% on average.}
    \label{fig:optim-breakdown-power}
\vspace{-9pt}
\end{minipage}
\end{figure}
\begin{table}[t]
\centering
\caption{Comparison between LEGO and related work with the same latency performance.}
\scalebox{1.0}{
\begin{tabular}{lccc}
\toprule
\multirow{2}{*}{Related Work}        & \multirow{2}{*}{\textbf{DSAGen}~\cite{weng2020dsagen}}      & \multirow{2}{*}{\textbf{TensorLib}{~\cite{jia2021tensorlib}}} & \multirow{2}{*}{}  \\
 &   &  \\ \midrule
LEGO & 2.6× power savings & 2.6× power savings                   \\
Improvement & 2.4× area savings  & 2.0× area savings   \\ 
\toprule
\multirow{2}{*}{Related Work}       & \textbf{AutoSA}~\cite{wang2021autosa} \textbf{based} & \multirow{2}{*}{\textbf{HLS-based SODA}{~\cite{agostini2022soda}}} 
\\
     &   \textbf{on Polyhedral Model} & \\ \midrule
LEGO &  6.5× FF savings & 32{×} power savings                \\
Improvement & 5.0× LUT savings & 14{×} speedup  \\ 
\bottomrule
\end{tabular}
}
\label{tab:comparison-related}
\vspace{-9pt}
\end{table}

\paragraph{Scalability} 
As discussed in {Section~\ref{sect:overview}}, the design can be scaled up either by increasing \#FUs in the lowest level or increasing \#PEs connected with L2 NoC at the higher level.
As shown in \tab{tab:scaleup}, we directly change the FU array size when \#FUs is below 1024 and scale the L2 NoC size for larger designs. The generation time is within 3 minutes, even for a large design with 16k FUs, which is far less than that of synthesis (typically a few hours). Moreover, scaling L2 NoC introduces $<$10\% overhead to the area and power and can be place-and-routing-friendly with a wormhole-based design.

\paragraph{Breakdown \& Ablation Study}
\fig{fig:hardware-breakdown}(a) demonstrates the area and power breakdown of different components in LEGO-generated design used in \fig{fig:savings-end-to-end}. On-chip buffers dominate the area cost with 86\% coverage, and FU Array, along with NoC, covers 83\% of on-chip power, while post-processing units only take up to 2\% area with 5\% power consumption.
\fig{fig:hardware-breakdown}(b) illustrates the end-to-end latency breakdown. Non-tensor operations ({\eg}, activation functions), are typically memory-bounded with little data reuse, making it hard to benefit from accelerators. However, the latency overhead brought by these operations on the post-processing units is less than 7.2\%.
For LLMs, the overhead of PPUs is also very low, ranging from 0.1\% (bs=1) to 1.4\% (bs=16) on LLaMA-7B. The overhead percentage is not expected to increase with future models with longer sequence lengths since both attention (processed by LEGO generated FU array) and activation function (hand-crafted PPU) scales at the same time with the token length.

As shown in \tab{tab:ablation-frontend},
different design goals may prefer different spatial dataflows:
LEGO-ICOCICOC offers higher performance, while LEGO-OHOWICOC provides better energy efficiency. By supporting the generation of arbitrary dataflow, LEGO can adapt to different design objectives. Furthermore, LEGO can fuse multiple spatial dataflows in a single design: the fused version (LEGO-MNICOC) achieves better performance on MobileNetV2 and Resnet50, and with the help of heuristic-based interconnection planning in Section~\ref{sect:frontend}, the overhead of fusion is minimized. Compared to single dataflow designs, our fused design achieves a better trade-off between performance and energy efficiency on multiple tasks.

\fig{fig:optim-breakdown-area} and \ref{fig:optim-breakdown-power} shows the contribution breakdown.
Reduction tree extraction and broadcast reduce the number of pipeline registers; each brings 15\% area saving on average. Power gating reduces the toggle rate on unused paths, saving 1.5\% power on average (9\% in Attention). Pin reusing saves the cost of adders and extra registers, reducing 5\% area and power on average.

We also evaluated the system overhead of LEGO by analyzing the instruction overhead. Among 7 benchmarks in~\ref{fig:savings-end-to-end}, the average \#cycles per instruction is over 2000, and the bandwidth requirement for instruction issue is between 0.05 and 0.13 GB/s, less than 1\% of the total DRAM bandwidth.

\paragraph{Comparison to Related Work} LEGO could achieve comparable results compared to handwritten designs (\tab{tab:compare-eyeriss-nvdla}), while DSAGen~\cite{weng2020dsagen} introduces 2.4× area and 2.6× power overhead (\tab{tab:comparison-related}).
Compared to TensorLib~\cite{jia2021tensorlib}, LEGO could reduce the flip-flop usage of the generated FU arrays by 70\% and combinational logic by 35\%, resulting in 2.6x power efficiency and 2.0× area efficiency (\tab{tab:comparison-related}). This is due to (a) the introduction of control flow analysis, which enables the sharing of memory control logic across FUs, and (b) the linear programming in the back end which takes the register usage as the objective function.

Compared to HLS-based SODA optimizer{~\cite{agostini2022soda}}, LEGO can generate FU array with similar hardware resource consumption with 14{×} performance and 32{×} energy efficiency on MobileNetV2. 
Compared to AutoSA~\cite{wang2021autosa}, LEGO saves up to 5.0× LUT and 6.5× FF resources (mainly the control logics in FUs and the data paths between them) when generating the same GEMM-IJ design on Xilinx Ultrascale+ FPGA with the same computation (\#DSPs) and memory resources (block rams). Table~{\ref{tab:comparison-soda}} and {\ref{tab:comparison-autosa}} provide more detailed comparison results.

Furthermore, LEGO can provide an accurate estimation of area/power/performance as feedback to DSE tools, and generate RTL for the design discovered by DSE tools. With the same resources as Eyeriss, using LEGO to generate the design searched by Timeloop~\cite{parashar2019timeloop} can reduce the power by 9\% while keeping the same latency performance. 

\begin{table}[t]
\centering
\caption{Comparison between LEGO and SODA toolchain (synthesis @ FreePDK 45nm with 500MHz)}
\label{tab:comparison-soda}
\addtolength{\tabcolsep}{-3pt}  
\scalebox{0.9}{
\begin{tabular}{p{40pt}|ccc|ccc}
\toprule
  & \multicolumn{3}{c|}{SODA+MLIR+Bambu~\cite{agostini2022soda}} & \multicolumn{3}{c}{LEGO (MNICOC-Tiny, 16 FUs)} \\ 
 NN Models & LeNet   & MobileNetV2   & Resnet50  & LeNet       & MobileNetV2       & Resnet50      \\ \midrule
Area (mm2)                                       & 0.67   & 0.75         & 0.41     & 0.945 & 0.945 & 0.945                        \\
GFLOPS                                           & 0.90   & 0.87         & 0.65     & 10.23       & 14.21             & 15.03         \\
GFLOPS/W                                         & 3.27    & 2.28          & 3.20      & 52.33       & 72.69             & 76.88        \\ \bottomrule
\end{tabular}}
\addtolength{\tabcolsep}{3pt}
\vspace{-9pt}
\end{table}
\begin{table}[t]
\centering
\caption{Comparison between LEGO and AutoSA on Xilinx Ultrascale+ U280 FPGA}
\label{tab:comparison-autosa}
\addtolength{\tabcolsep}{-3pt}  
\scalebox{0.75}{
\begin{tabular}{p{25pt}|ccc|ccc}
\toprule
Tensor  & \multicolumn{3}{c|}{AutoSA~\cite{wang2021autosa}} & \multicolumn{3}{c}{\textbf{LEGO}} \\ 
Kernels & GEMM-IJ   & Conv2d-OCOH   & MTTKRP-IJ  & GEMM-IJ       & Conv2d-OCOH       & MTTKRP-IJ      \\ \midrule
FF                                       & 25.4K   &  108K        & 96.0K     & 3.9K & 4.9K & 4.9K                        \\
LUT                                           & 23.9K   & 120K         & 92.4K    & 4.8K       & 4.2K             & 4.7K        \\\bottomrule
\end{tabular}}
\addtolength{\tabcolsep}{3pt}  
\vspace{-9pt}
\end{table}

\section{Related Work}
\label{sect:related} 
\paragraph{DSE Frameworks} 
Prior design space exploration (DSE) frameworks for spatial architectures mostly focus on neural networks.
TimeLoop~\cite{parashar2019timeloop} uses a compute-centric notation while 
MAESTRO~\cite{kwon2019understanding} proposes a data-centric notation for dataflow representation. 
TENET~\cite{lu2021tenet} applies relation-centric notation to extend the scope to tensor applications.
MAGNET~\cite{venkatesan2019magnet} leverages Timeloop for automatic design space exploration with parameterized hardware templates.
NAAS~\cite{lin2021naas} exploits MAESTRO to further explore the co-design of neural network architecture. LEGO targets automatic RTL generation and can be used in serial with these works: \textit{LEGO generates the RTL of the design discovered by the DSE tools.} 

\paragraph{Accelerator Generators} Template-based spatial architecture generators, including Gemmini~\cite{genc2021gemmini}, DNA~\cite{zhang2020dna}, DNNWeaver~\cite{sharma2016dnnweaver}, and MAGNET~\cite{venkatesan2019magnet}, require handwritten templates and apply top-down parameter configuration.
TensorLib{~\cite{jia2021tensorlib}}, EMS{~\cite{jia2022ems}} and LEGO build the FU array in a bottom-up way by connecting FUs based on the data reuse analysis.
More comparisons against TensorLib are discussed in \sect{sect:representation-differences}, \sect{sect:frontend-interconnection-analysis}.
Without templates, DSAGen~\cite{weng2020dsagen} is a CGRA-based generator that targets high reconfigurability. SODA toolchain{~\cite{agostini2022soda}}, incorporating MLIR{~\cite{lattner2021mlir}} and Bambu{~\cite{ferrandi2021bambu}}, provides an open-sourced end-to-end automatic generation pipeline from Python-based NN to hardware RTL. DSAGen uses flexible switches to distribute data and thus natively supports sparse operations, while SODA exploits HLS for hardware generation, and thus most optimizations are general and HLS-oriented. In contrast, LEGO targets performance on ASICs, directly generates and optimizes the hardware by maximizing data reuse and minimizing the data path cost. 

\paragraph{Polyhedral Model} 
The polyhedral model is a widely used framework for analyzing and optimizing loop nests. PolySA~\cite{cong2018polysa} and its successor AutoSA~\cite{wang2021autosa} use the polyhedral model to generate systolic arrays automatically.
More detailed differences are discussed in \sect{sect:representation-differences}.

\section{Conclusion}
\label{sect:conclusion}
We provide an automatic design generation methodology, LEGO for spatial accelerators without templates. LEGO introduces a high-level relation-centric hardware representation, builds the FU-level interconnections via graph algorithms, and optimizes the primitive-level graph via linear-programming-based algorithms.
LEGO enables a larger hardware design space, and more efficient development of spatial accelerators, allowing for greater potential for customization for diverse tensor operations in emerging applications.

\section*{Acknowledgements}

We thank MIT AI Hardware Program, National Science Foundation, MIT-IBM Watson AI Lab, and Amazon for supporting this research. 

\bibliographystyle{IEEEtranS}
\bibliography{refs}

\begin{thebibliography}{10}
\providecommand{\url}[1]{#1}
\csname url@samestyle\endcsname
\providecommand{\newblock}{\relax}
\providecommand{\bibinfo}[2]{#2}
\providecommand{\BIBentrySTDinterwordspacing}{\spaceskip=0pt\relax}
\providecommand{\BIBentryALTinterwordstretchfactor}{4}
\providecommand{\BIBentryALTinterwordspacing}{\spaceskip=\fontdimen2\font plus
\BIBentryALTinterwordstretchfactor\fontdimen3\font minus \fontdimen4\font\relax}
\providecommand{\BIBforeignlanguage}[2]{{%
\expandafter\ifx\csname l@#1\endcsname\relax
\typeout{** WARNING: IEEEtranS.bst: No hyphenation pattern has been}%
\typeout{** loaded for the language `#1'. Using the pattern for}%
\typeout{** the default language instead.}%
\else
\language=\csname l@#1\endcsname
\fi
#2}}
\providecommand{\BIBdecl}{\relax}
\BIBdecl

\bibitem{agostini2022soda}
N.~B. Agostini, A.~Limaye, M.~Minutoli, V.~G. Castellana, J.~Manzano, A.~Tumeo, S.~Curzel, and F.~Ferrandi, ``Soda synthesizer: an open-source, multi-level, modular, extensible compiler from high-level frameworks to silicon,'' in \emph{Proceedings of the 41st IEEE/ACM International Conference on Computer-Aided Design}, 2022, pp. 1--7.

\bibitem{baltus1993efficient}
D.~G. Baltus and J.~Allen, ``Efficient exploration of nonuniform space-time transformations for optimal systolic array synthesis,'' in \emph{Proceedings of International Conference on Application Specific Array Processors (ASAP'93)}.\hskip 1em plus 0.5em minus 0.4em\relax IEEE, 1993, pp. 428--441.

\bibitem{chen2016eyeriss}
Y.-H. Chen, T.~Krishna, J.~S. Emer, and V.~Sze, ``Eyeriss: An energy-efficient reconfigurable accelerator for deep convolutional neural networks,'' \emph{IEEE journal of solid-state circuits}, vol.~52, no.~1, pp. 127--138, 2016.

\bibitem{choy20194d}
C.~Choy, J.~Gwak, and S.~Savarese, ``4d spatio-temporal convnets: Minkowski convolutional neural networks,'' in \emph{Proceedings of the IEEE/CVF conference on computer vision and pattern recognition}, 2019, pp. 3075--3084.

\bibitem{cong2018polysa}
J.~Cong and J.~Wang, ``Polysa: Polyhedral-based systolic array auto-compilation,'' in \emph{2018 IEEE/ACM International Conference on Computer-Aided Design (ICCAD)}.\hskip 1em plus 0.5em minus 0.4em\relax IEEE, 2018, pp. 1--8.

\bibitem{dai2021coatnet}
Z.~Dai, H.~Liu, Q.~V. Le, and M.~Tan, ``Coatnet: Marrying convolution and attention for all data sizes,'' \emph{Advances in Neural Information Processing Systems}, vol.~34, pp. 3965--3977, 2021.

\bibitem{devlin2018bert}
J.~Devlin, M.-W. Chang, K.~Lee, and K.~Toutanova, ``Bert: Pre-training of deep bidirectional transformers for language understanding,'' \emph{arXiv preprint arXiv:1810.04805}, 2018.

\bibitem{du2015shidiannao}
Z.~Du, R.~Fasthuber, T.~Chen, P.~Ienne, L.~Li, T.~Luo, X.~Feng, Y.~Chen, and O.~Temam, ``Shidiannao: Shifting vision processing closer to the sensor,'' in \emph{Proceedings of the 42nd Annual International Symposium on Computer Architecture}, 2015, pp. 92--104.

\bibitem{ferrandi2021bambu}
F.~Ferrandi, V.~G. Castellana, S.~Curzel, P.~Fezzardi, M.~Fiorito, M.~Lattuada, M.~Minutoli, C.~Pilato, and A.~Tumeo, ``Bambu: an open-source research framework for the high-level synthesis of complex applications,'' in \emph{2021 58th ACM/IEEE Design Automation Conference (DAC)}.\hskip 1em plus 0.5em minus 0.4em\relax IEEE, 2021, pp. 1327--1330.

\bibitem{genc2021gemmini}
H.~Genc, S.~Kim, A.~Amid, A.~Haj-Ali, V.~Iyer, P.~Prakash, J.~Zhao, D.~Grubb, H.~Liew, H.~Mao \emph{et~al.}, ``Gemmini: Enabling systematic deep-learning architecture evaluation via full-stack integration,'' in \emph{2021 58th ACM/IEEE Design Automation Conference (DAC)}.\hskip 1em plus 0.5em minus 0.4em\relax IEEE, 2021, pp. 769--774.

\bibitem{copilot}
GitHub, ``{GitHub Copilot: AI pair programmer},'' \url{https://copilot.github.com/}, 2021.

\bibitem{ham2020a3}
T.~J. Ham, S.~J. Jung, S.~Kim, Y.~H. Oh, Y.~Park, Y.~Song, J.-H. Park, S.~Lee, K.~Park, J.~W. Lee \emph{et~al.}, ``A\^{3}: Accelerating attention mechanisms in neural networks with approximation,'' in \emph{2020 IEEE International Symposium on High Performance Computer Architecture (HPCA)}.\hskip 1em plus 0.5em minus 0.4em\relax IEEE, 2020, pp. 328--341.

\bibitem{he2016deep}
K.~He, X.~Zhang, S.~Ren, and J.~Sun, ``Deep residual learning for image recognition,'' in \emph{Proceedings of the IEEE conference on computer vision and pattern recognition}, 2016, pp. 770--778.

\bibitem{ho2020denoising}
J.~Ho, A.~Jain, and P.~Abbeel, ``Denoising diffusion probabilistic models,'' \emph{Advances in Neural Information Processing Systems}, vol.~33, pp. 6840--6851, 2020.

\bibitem{huangfu2018parallelizing}
Q.~Huangfu and J.~J. Hall, ``Parallelizing the dual revised simplex method,'' \emph{Mathematical Programming Computation}, vol.~10, no.~1, pp. 119--142, 2018.

\bibitem{jia2021tensorlib}
L.~Jia, Z.~Luo, L.~Lu, and Y.~Liang, ``Tensorlib: A spatial accelerator generation framework for tensor algebra,'' in \emph{2021 58th ACM/IEEE Design Automation Conference (DAC)}.\hskip 1em plus 0.5em minus 0.4em\relax IEEE, 2021, pp. 865--870.

\bibitem{jia2022ems}
L.~Jia, Y.~Wang, J.~Leng, and Y.~Liang, ``Ems: Efficient memory subsystem synthesis for spatial accelerators,'' in \emph{Proceedings of the 59th ACM/IEEE Design Automation Conference}, 2022, pp. 67--72.

\bibitem{jouppi2017datacenter}
N.~P. Jouppi, C.~Young, N.~Patil, D.~Patterson, G.~Agrawal, R.~Bajwa, S.~Bates, S.~Bhatia, N.~Boden, A.~Borchers \emph{et~al.}, ``In-datacenter performance analysis of a tensor processing unit,'' in \emph{Proceedings of the 44th annual international symposium on computer architecture}, 2017, pp. 1--12.

\bibitem{krizhevsky2017imagenet}
A.~Krizhevsky, I.~Sutskever, and G.~E. Hinton, ``Imagenet classification with deep convolutional neural networks,'' \emph{Communications of the ACM}, vol.~60, no.~6, pp. 84--90, 2017.

\bibitem{kwon2019understanding}
H.~Kwon, P.~Chatarasi, M.~Pellauer, A.~Parashar, V.~Sarkar, and T.~Krishna, ``Understanding reuse, performance, and hardware cost of dnn dataflow: A data-centric approach,'' in \emph{Proceedings of the 52nd Annual IEEE/ACM International Symposium on Microarchitecture}, 2019, pp. 754--768.

\bibitem{kwon2018maeri}
H.~Kwon, A.~Samajdar, and T.~Krishna, ``Maeri: Enabling flexible dataflow mapping over dnn accelerators via reconfigurable interconnects,'' \emph{ACM SIGPLAN Notices}, vol.~53, no.~2, pp. 461--475, 2018.

\bibitem{lattner2021mlir}
C.~Lattner, M.~Amini, U.~Bondhugula, A.~Cohen, A.~Davis, J.~Pienaar, R.~Riddle, T.~Shpeisman, N.~Vasilache, and O.~Zinenko, ``Mlir: Scaling compiler infrastructure for domain specific computation,'' in \emph{2021 IEEE/ACM International Symposium on Code Generation and Optimization (CGO)}.\hskip 1em plus 0.5em minus 0.4em\relax IEEE, 2021, pp. 2--14.

\bibitem{lin2021naas}
Y.~Lin, M.~Yang, and S.~Han, ``Naas: Neural accelerator architecture search,'' in \emph{2021 58th ACM/IEEE Design Automation Conference (DAC)}.\hskip 1em plus 0.5em minus 0.4em\relax IEEE, 2021, pp. 1051--1056.

\bibitem{lin2021pointacc}
Y.~Lin, Z.~Zhang, H.~Tang, H.~Wang, and S.~Han, ``Pointacc: Efficient point cloud accelerator,'' in \emph{MICRO-54: 54th Annual IEEE/ACM International Symposium on Microarchitecture}, 2021, pp. 449--461.

\bibitem{lu2021tenet}
L.~Lu, N.~Guan, Y.~Wang, L.~Jia, Z.~Luo, J.~Yin, J.~Cong, and Y.~Liang, ``Tenet: A framework for modeling tensor dataflow based on relation-centric notation,'' in \emph{2021 ACM/IEEE 48th Annual International Symposium on Computer Architecture (ISCA)}.\hskip 1em plus 0.5em minus 0.4em\relax IEEE, 2021, pp. 720--733.

\bibitem{lu2021sanger}
L.~Lu, Y.~Jin, H.~Bi, Z.~Luo, P.~Li, T.~Wang, and Y.~Liang, ``Sanger: A co-design framework for enabling sparse attention using reconfigurable architecture,'' in \emph{MICRO-54: 54th Annual IEEE/ACM International Symposium on Microarchitecture}, 2021, pp. 977--991.

\bibitem{muralimanohar2009cacti}
N.~Muralimanohar, R.~Balasubramonian, and N.~P. Jouppi, ``Cacti 6.0: A tool to model large caches,'' \emph{HP laboratories}, vol.~27, p.~28, 2009.

\bibitem{chatgpt}
OpenAI, ``{ChatGPT: a large language model trained on the GPT-3.5 architecture},'' \url{https://openai.com/}, 2021.

\bibitem{gpt4}
OpenAI, ``Gpt-4 technical report,'' \emph{arXiv preprint arXiv:2303.08774}, 2023.

\bibitem{parashar2019timeloop}
A.~Parashar, P.~Raina, Y.~S. Shao, Y.-H. Chen, V.~A. Ying, A.~Mukkara, R.~Venkatesan, B.~Khailany, S.~W. Keckler, and J.~Emer, ``Timeloop: A systematic approach to dnn accelerator evaluation,'' in \emph{2019 IEEE international symposium on performance analysis of systems and software (ISPASS)}.\hskip 1em plus 0.5em minus 0.4em\relax IEEE, 2019, pp. 304--315.

\bibitem{radford2019language}
A.~Radford, J.~Wu, R.~Child, D.~Luan, D.~Amodei, I.~Sutskever \emph{et~al.}, ``Language models are unsupervised multitask learners,'' \emph{OpenAI blog}, vol.~1, no.~8, p.~9, 2019.

\bibitem{rombach2022high}
R.~Rombach, A.~Blattmann, D.~Lorenz, P.~Esser, and B.~Ommer, ``High-resolution image synthesis with latent diffusion models,'' in \emph{Proceedings of the IEEE/CVF Conference on Computer Vision and Pattern Recognition}, 2022, pp. 10\,684--10\,695.

\bibitem{sandler2018mobilenetv2}
M.~Sandler, A.~Howard, M.~Zhu, A.~Zhmoginov, and L.-C. Chen, ``Mobilenetv2: Inverted residuals and linear bottlenecks,'' in \emph{Proceedings of the IEEE conference on computer vision and pattern recognition}, 2018, pp. 4510--4520.

\bibitem{sharma2016dnnweaver}
H.~Sharma, J.~Park, E.~Amaro, B.~Thwaites, P.~Kotha, A.~Gupta, J.~K. Kim, A.~Mishra, and H.~Esmaeilzadeh, ``Dnnweaver: From high-level deep network models to fpga acceleration,'' in \emph{the Workshop on Cognitive Architectures}, 2016.

\bibitem{sharma2018bit}
H.~Sharma, J.~Park, N.~Suda, L.~Lai, B.~Chau, V.~Chandra, and H.~Esmaeilzadeh, ``Bit fusion: Bit-level dynamically composable architecture for accelerating deep neural network,'' in \emph{2018 ACM/IEEE 45th Annual International Symposium on Computer Architecture (ISCA)}.\hskip 1em plus 0.5em minus 0.4em\relax IEEE, 2018, pp. 764--775.

\bibitem{tan2021efficientnetv2}
M.~Tan and Q.~Le, ``Efficientnetv2: Smaller models and faster training,'' in \emph{International conference on machine learning}.\hskip 1em plus 0.5em minus 0.4em\relax PMLR, 2021, pp. 10\,096--10\,106.

\bibitem{tarjan1977finding}
R.~E. Tarjan, ``Finding optimum branchings,'' \emph{Networks}, vol.~7, no.~1, pp. 25--35, 1977.

\bibitem{touvron2023llama}
H.~Touvron, T.~Lavril, G.~Izacard, X.~Martinet, M.-A. Lachaux, T.~Lacroix, B.~Rozi{\`e}re, N.~Goyal, E.~Hambro, F.~Azhar \emph{et~al.}, ``Llama: Open and efficient foundation language models,'' \emph{arXiv preprint arXiv:2302.13971}, 2023.

\bibitem{vaswani2017attention}
A.~Vaswani, N.~Shazeer, N.~Parmar, J.~Uszkoreit, L.~Jones, A.~N. Gomez, {\L}.~Kaiser, and I.~Polosukhin, ``Attention is all you need,'' \emph{Advances in neural information processing systems}, vol.~30, 2017.

\bibitem{venkatesan2019magnet}
R.~Venkatesan, Y.~S. Shao, M.~Wang, J.~Clemons, S.~Dai, M.~Fojtik, B.~Keller, A.~Klinefelter, N.~Pinckney, P.~Raina \emph{et~al.}, ``Magnet: A modular accelerator generator for neural networks,'' in \emph{2019 IEEE/ACM International Conference on Computer-Aided Design (ICCAD)}.\hskip 1em plus 0.5em minus 0.4em\relax IEEE, 2019, pp. 1--8.

\bibitem{wang2021spatten}
H.~Wang, Z.~Zhang, and S.~Han, ``Spatten: Efficient sparse attention architecture with cascade token and head pruning,'' in \emph{2021 IEEE International Symposium on High-Performance Computer Architecture (HPCA)}.\hskip 1em plus 0.5em minus 0.4em\relax IEEE, 2021, pp. 97--110.

\bibitem{wang2021autosa}
J.~Wang, L.~Guo, and J.~Cong, ``Autosa: A polyhedral compiler for high-performance systolic arrays on fpga,'' in \emph{The 2021 ACM/SIGDA International Symposium on Field-Programmable Gate Arrays}, 2021, pp. 93--104.

\bibitem{weng2020dsagen}
J.~Weng, S.~Liu, V.~Dadu, Z.~Wang, P.~Shah, and T.~Nowatzki, ``Dsagen: Synthesizing programmable spatial accelerators,'' in \emph{2020 ACM/IEEE 47th Annual International Symposium on Computer Architecture (ISCA)}.\hskip 1em plus 0.5em minus 0.4em\relax IEEE, 2020, pp. 268--281.

\bibitem{wu201316nm}
S.-Y. Wu, C.~Y. Lin, M.~Chiang, J.~Liaw, J.~Cheng, S.~Yang, M.~Liang, T.~Miyashita, C.~Tsai, B.~Hsu \emph{et~al.}, ``A 16nm finfet cmos technology for mobile soc and computing applications,'' in \emph{2013 IEEE International Electron Devices Meeting}.\hskip 1em plus 0.5em minus 0.4em\relax IEEE, 2013, pp. 9--1.

\bibitem{zhang2020dna}
Y.~Zhang, Y.~Fu, W.~Jiang, C.~Li, H.~You, M.~Li, V.~Chandra, and Y.~Lin, ``Dna: Differentiable network-accelerator co-search,'' \emph{arXiv preprint arXiv:2010.14778}, 2020.

\bibitem{zhou2018research}
G.~Zhou, J.~Zhou, and H.~Lin, ``Research on nvidia deep learning accelerator,'' in \emph{2018 12th IEEE International Conference on Anti-counterfeiting, Security, and Identification (ASID)}.\hskip 1em plus 0.5em minus 0.4em\relax IEEE, 2018, pp. 192--195.

\end{thebibliography}

\end{document}